\documentclass[review,authoryear]{elsarticle}

\usepackage{lineno,hyperref}
\usepackage{booktabs}
\usepackage{xcolor}[table]
\usepackage{diagbox}
\usepackage{color}
\usepackage{colortbl}
\usepackage{multirow}
\usepackage{subcaption}
\usepackage[inline]{enumitem}
\usepackage{soul}
\usepackage{amsmath,amssymb}

\usepackage[utf8x]{inputenc}	

\journal{Advances in Space Research}

\begin{document}

\begin{frontmatter}

\title{Preliminary mission profile of Hera's Milani CubeSat}

\author{Fabio Ferrari\corref{cor}}
\ead{fabio1.ferrari@polimi.it}
\author{Vittorio Franzese}
\ead{vittorio.franzese@polimi.it}
\author{Mattia Pugliatti}
\ead{mattia.pugliatti@polimi.it}
\author{Carmine Giordano}
\ead{carmine.giordano@polimi.it}
\author{Francesco Topputo}
\ead{francesco.topputo@polimi.it}
\address{Department of Aerospace Science and Technology, Politecnico~di~Milano, Via~La~Masa~34, 20156, Milan, Italy}
\cortext[cor]{Corresponding author}

\begin{abstract} 
    CubeSats offer a flexible and low-cost option to increase the scientific and technological return of small-body exploration missions. ESA's Hera mission, the European component of the Asteroid Impact and Deflection Assessment (AIDA) international collaboration, plans on deploying two CubeSats in the proximity of binary system 65803~Didymos, after arrival in 2027.
    In this work, we discuss the feasibility and preliminary mission profile of Hera's Milani CubeSat. The CubeSat mission is designed to achieve both scientific and technological objectives. We identify the design challenges and discuss design criteria to find suitable solutions in terms of mission analysis, operational trajectories, and Guidance, Navigation, \& Control (GNC) design. We present initial trajectories and GNC baseline, as a result of trade-off analyses. We assess the feasibility of the Milani CubeSat mission and provide a preliminary solution to cover the operational mission profile of Milani in the close-proximity of Didymos system.
\end{abstract}


\end{frontmatter}

\clearpage
\section{Introduction}
\label{sec:1-intro}
    Motivated by a great scientific interest and large accessibility to space missions, small celestial bodies represent the current frontier of space exploration. The close-proximity exploration of asteroids and comets entered a new era in 2000, as the NEAR-Shoemaker spacecraft entered into orbit around asteroid 433 Eros, achieving the first rendezvous with a small celestial body~\citep{NEARShoe}. An important milestone was marked more recently, by ESA's Rosetta~\citep{RosettaPhilaeMASCOT} and later by JAXA's Hayabusa 2~\citep{VanWal2018,Tsuda2020} missions, which were able to release small probes and SmallSats on the surface and/or in the close-proximity of such objects. In this context, CubeSats offer a flexible and low-cost option to increase the scientific and technological return of small-body exploration missions. Although many concepts of interplanetary CubeSats have been studied and/or are planned for future missions~\citep{Walker2018,Franzese2019,Speretta2019,topputo2020envelop}, very few of them have flown so far~\citep{Freeman2020,Lockett2020}, and no CubeSat has ever flown in the proximity of a small-body yet. 
    
    ESA's Hera mission~\citep{Michel2018}, the European component of the Asteroid Impact and Deflection Assessment (AIDA) international collaboration~\citep{AIDA_AA}, plans on deploying two CubeSats in the proximity of binary system 65803~Didymos, after arrival in 2027. The idea is supported by several SmallSat/CubeSat studies that have been performed in the past few years with reference to Hera (formerly known as AIM), to study the feasibility of deploying a small lander~\citep{Ferrari2018b,AIM_AAS16_MEX,Lange2018,Celik2019} and possibly CubeSats~\citep{Kohout2018,LasagniManghi2018,Perez2018} in the close proximity of Didymos. These were anticipated by detailed studies of the dynamical environment near binary asteroid systems~\citep{Tardivel2013a,Tardivel2016,Ferrari2016,FerrariAstAAS14,Celik2017}, some with direct application to the Didymos case study~\citep{DellElce2017,Capannolo2019,CapannoloFerrariIAC16}. Hera is currently in its phase-B design, on-track to be launched in 2024. Its payloads have been recently consolidated and, as mentioned, it will carry two 6U CubeSats, to be deployed after rendezvous with Didymos. These are called Juventas~\citep{Karatekin2019} and Milani. Milani has been officially confirmed on mid 2020 and is currently in its phase-A design. Its flight readiness does reflect Hera schedule, foreseen in late 2024.
    
    In this work, we discuss the feasibility and preliminary mission profile of Hera's Milani CubeSat. The Milani mission is designed to achieve both scientific and technological objectives. The first include the global mapping of Didymos (primary) and Dimorphos (secondary) asteroids. This involves determining the composition difference between the bodies and the study of their surface properties. In the context of the AIDA international collaboration, Milani will help at determining the composition of the ejecta created after the impact of DART spacecraft~\citep{Cheng2018} in 2022 on the surface of Dimorphos, and their fallbacks on the surface of the asteroids. In terms of the technological objectives, the mission will be among the first to operate CubeSat technologies in deep space and the first to use Inter-Satellite Link (ISL) network with the mother spacecraft in such challenging context.

    Milani is a 6U CubeSat with maneuvering capabilities both in terms of translational and attitude motion (full 6-DOF). After release it will perform autonomous operations to reach its operational trajectories towards the fulfilment of its mission goals. The main payload aboard Milani is the ASPECT hyperspectral camera~\citep{Kohout2018}. The characteristics of the ASPECT camera are reported in Table~\ref{tab:ASPECT}. The payload has three channels, which are the Visible (VIS), the Near-Infrared (NIR), and the Short Wavelength Infrared (SWIR). The three channels have different characteristics, especially in terms of field-of-view and sensor size. These quantities affect the allowed orbital ranges to the targets for scientific observations. The visible channel has the largest field of view and a 1~Mpx sensor size. The NIR is narrower than the VIS channel, while the SWIR is a mono-pixel circular FOV channel. In addition to ASPECT, which is used for scientific imaging only, Milani hosts two additional cameras, which are used for optical navigation. These cameras have not been selected yet: the properties and performance assumed for navigation cameras are reported in Section~\ref{subsec:opnav}. As a seconday payload, Milani hosts the VISTA thermogravimeter~\citep{Dirri2018}, which will collect information on the dust environment near the asteroids. It is operated in background during the operational phases of the Milani mission: it does not impose any further requirement and does not require a specific tayloring of the mission profile.

    \begin{table}[t]
        \centering
        \caption{Characteristics of the ASPECT hyperspectral camera.}
        \begin{tabular}{lcccc}
            \toprule
            \toprule
            \textbf{Parameter} & \textbf{Unit} & \textbf{VIS} & \textbf{NIR} & \textbf{SWIR}   \\
            \midrule
            Field-Of-View & deg & 10 x 10 & 6.68 x 5.36 & 5 \\
            Sensor Size & pix & 1024 x 1024 & 640 x 512 & 1 \\
            Pixel Size & $\mu$m & 5.5 x 5.5 & 15 x 15 & 1000 \\
            Focal Length & mm & 32.3 & 81.7 & 11.7 \\
            F-number & - & 3.3 & 6.0 & 0.9 \\
            \bottomrule
        \end{tabular}
        \label{tab:ASPECT}
    \end{table}

    After release in the proximity of Didymos binary asteroid, Milani independently makes its own way to its operational trajectory, optimized to observe the asteroid system. In this work, we consider the CubeSat limitations given by miniaturized components, and address the challenges in deriving an optimal mission profile which meets the mission and scientific objectives. The GNC architecture and design that allow to achieve such a mission are also shown. It is worth stressing that the initial results presented in this paper have been obtained in response to the relevant ESA call. As such, the preliminary profile here shown will be revised as both mission and spacecraft design advance in later stages.

    The paper is organized as follows. Section~\ref{sec:2-req} reports the design constraints of the mission given by the environment and the payload. Section~\ref{sec:3-profile} shows the mission profile with a detailed explanation of the mission phases. Then, the GNC architecture and preliminary design are shown in Section~\ref{sec:4-gnc}. The operational analysis of the payload is reported in Section~\ref{sec:5-payload}. Finally, conclusions and a summary of the results are given in Section~\ref{sec:6-conclusion}.

\section{Design constraints} 
\label{sec:2-req}
    The constraints affecting the design of the mission and the CubeSat platform are presented in this Section. The Didymos binary system is composed of two asteroids: the primary (Didymos, also called D1 in the following) and its satellite (Dimorphos, also called the secondary or D2 in the following). The primary body has an estimated diameter of 780~m and an assumed albedo of 0.15. The secondary has an estimated diameter of 170~m and an assumed albedo of 0.15. Current estimates consider a 10~\% size error for both asteroids. The requirements applicable to the Milani CubeSat derive directly from both the mission objectives and the ASPECT payload. To achieve the mission objectives with ASPECT, the payload team has determined the following two main conditions:
    1) To map both asteroids with a ground resolution lower than 2~m/pixel with images from both the VIS and the NIR channels; 2) To observe the asteroids with a phase angle (Sun-asteroid-CubeSat angle) between 5 and 25~degrees. In order to ease the operations of Milani, Dimorphos is required to be imaged within a single picture, always inside the FOV of ASPECT, while the primary can be imaged as a 4-images mosaic.
    These values are summarized in Table~\ref{tab:DidymosSystem}, where $d$ is the equivalent diameter of the asteroids, $m_d$ the associated margin, $p_v$ the albedo, GSD the ground sampling distance, $\alpha$ the phase angle, and $\theta/$FOV the ratio between the apparent dimension of the asteroid to the ASPECT field-of-view. Note that the limiting value for the FOV is given by the NIR channel, since it is narrower than the VIS channel. The SWIR channel is not considered here because it is a monopixel channel.

    \begin{table}[t]
        \centering
        \caption{Didymos system characteristics and imaging constraints.}
        \begin{tabular}{lcccccc}
            \toprule
            \toprule
            \textbf{Asteroid} & $d$ & $m_d$ & $p_v$ & GSD & $\alpha$ & $\theta$/FOV\\
            \midrule
            Didymos & 780 m & 10 \% & 0.15 & $\leq$ 2 m/pix & 5 -- 25 deg & $\leq$ 4 \\
            Dimorphos & 170 m & 10 \% & 0.15 & $\leq$ 2 m/pix & 5 -- 25 deg & $\leq$ 1 \\
            \bottomrule
        \end{tabular}
        \label{tab:DidymosSystem}
    \end{table}

With the values in Table~\ref{tab:DidymosSystem}, it is easy to verify that Dimorphos fills the FOV of the VIS channel at 1069~m of distance and the FOV of the NIR channel at 1997~m of distance, while Didymos fills a 2$\times$2 image mosaic using the VIS channel at a distance of 2433 m and the NIR channel at 4572 m. Thus, the value of 4572~m is considered as the lower bound on the range constraints to the binary system. The maximum allowable ranges to meet the 2~m/pix imaging GSD are 11700~m for the VIS channel and 10940~m for the NIR channel. Thus, the upper bound on the relative range between the binary system and Milani is 10940~m. The communication link between Milani and the mother spacecraft through the ISL is assumed to be omnidirectional and available up to 60~km of relative distance between the two spacecraft.
    
    All in all, the operations in the nominal orbit shall be performed at ranges from Didymos within 4572--10940~m and phase angles within 5 and 25~degrees, as summarized in Table~\ref{tab:Constraints}, within a 60~km range from Hera mothercraft.

    \begin{table}[ht]
        \centering
        \caption{Mission design constraints.}
        \begin{tabular}{lcccc}
            \toprule
            \toprule
            \textbf{Asteroid} & Min Range & Max Range & Min Phase Angle & Max Phase Angle \\
            \midrule
            Didymos & 4572 m & 10940 m & 5 deg & 25 deg  \\
            \bottomrule
        \end{tabular}
        \label{tab:Constraints}
    \end{table}

\section{Mission profile} 
\label{sec:3-profile}
    This section presents the preliminary mission profile for Hera's Milani CubeSat, from its release in the proximity of Didymos system, up to its decommisioning. We discuss the design strategy to provide efficient release of Milani from Hera spacecraft and detail the operational trajectories during the CubeSat science phase. End-of-life options are proposed. The feasibility of the mission profile is eventually discussed. The overall design is compliant with constraints outlined in Section~\ref{sec:2-req}. The dynamics of the CubeSat near the Didymos system are studied using a high-fidelity model, which includes the heliocentric motion of Didymos system barycenter, the binary motion of the asteroids (provided by high-order integration in ESA's Hera Mission kernels\footnote{\url{https://www.cosmos.esa.int/web/spice/data}, Last accessed: June 2020}) and relevant perturbations, such as the effects of the non-sphericity of asteroids (through a polyhedral model of Didymos and ellipsoidal model of Dimorphos) and Solar Radiation Pressure (SRP). 
    
    \subsection{Release strategy} 
    \label{sub:3.1-release}
        The Hera spacecraft plans on releasing Milani after arrival at Didymos system, in late spring~2027. In this work, we set the release date to May~1st, 2027. At this time the Didymos--Sun distance is 1.5~AU and increases up to 2.0~AU during the expected operational life of the CubeSat. The selection of Milani operational dates is important to characterize the dynamical environment in the proximity of Didymos, as it affects the magnitude of the SRP, which is relevant. On the other hand, the release date does not affect the operations of Milani in terms of orbital determination and illumination conditions. The relative geometry of Hera trajectories and release point is set in terms of Milani/Hera, Didymos and Sun position. At any epoch, the release will be performed when Hera is between the binary system and the Sun direction, in order to have Didymos visible on its day side.
        Milani's release conditions are carefully selected after an extensive analysis, involving the investigation of different release points on Hera's Early and Detailed Characterization Phases trajectories~\footnote{see Hera Mission kernels documentation at~\url{https://www.cosmos.esa.int/web/spice/data}}, the analysis of geometrical constraints at release, the dynamics of the CubeSat after release relative to Hera mothercraft and Didymos system. In particular, the following criteria have been considered in terms of release geometry and release trajectory:
        \begin{itemize}
            \item Minimum angle between CubeSat release velocity direction and CubeSat-Sun direction of 45~deg;
            \item Minimum safety factor C of the release trajectory of 0.4.
        \end{itemize}
        The safety factor C measures the energy of the trajectory (C$>$0 for hyperbolas), and is defined as
        \begin{equation}
            v_{\textrm{peri}}=(1+C)v_{\textrm{parab}}
        \end{equation}
        with $v_{\textrm{peri}}$ and $v_{\textrm{parab}}$ defined as the velocity at pericenter on the release trajectory and the parabolic velocity at the release point, respectively. These conditions derive from the mother spacecraft requirements in terms of safety and illumination conditions. In addition, the release trajectory was selected to guarantee consistent relative motion with respect to Hera spacecraft and Didymos system for up to 7~days after release, in terms of:
        \begin{itemize}
            \item Maximum CubeSat--Hera distance of~60 km, to ensure communication with the mothercraft;
            \item Maximum phase angle of Didymos system (Sun-Didymos-CubeSat angle) of 90~deg, to ensure dayside visibility of the asteroids;
        \end{itemize}
        Finally, we analysed the robustness to release uncertainties, related to both the deployer mechanism and Hera's GNC accuracy. These are quantified as:
        \begin{itemize}
            \item Release velocity in the range 0-5~cm/s ($\pm$1~cm/s, 3$\sigma$~value)
            \item Release direction accuracy of the deployer mechanism within 5~deg (3$\sigma$~value)
            \item Release orientation accuracy of Hera spacecraft within 0.5~deg (3$\sigma$~value)
        \end{itemize}
        The analysis performed is aimed at finding suitable release directions. In this case, pointing and release direction errors are the most important sources of errors. Position and velocity errors of Hera play a minor role in detecting preferred directions and, at the current stage they can be neglected. These will be added at a later refinement of the design.
        Release points are chosen among Hera spacecraft trajectories, which are retrieved from ESA's Hera Mission kernels.
        
\begin{figure}[t]
          	\centering
        	\includegraphics[width=\textwidth]{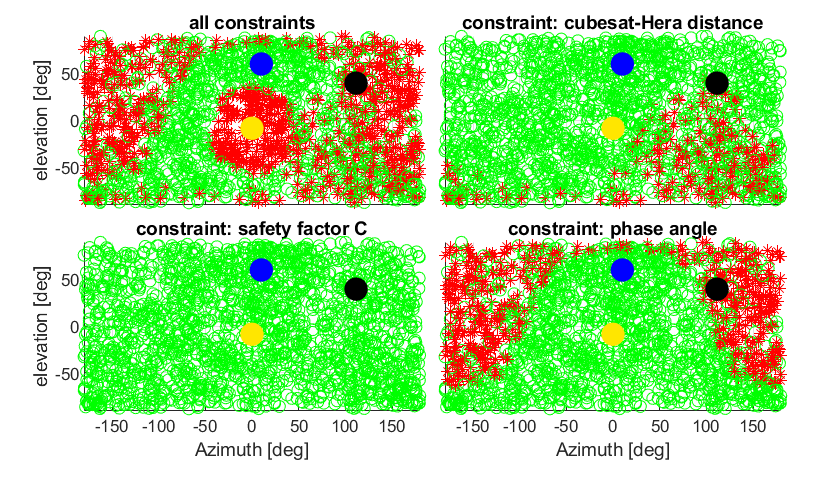}
           	\caption{Release options in the Didymos equatorial plane. Green and red markers indicate compliant and non-compliant solutions, respectively. The upper-left picture is the composite of the other three pictures, plus the addition of geometrical constraints at release (forbidden $\pm$45 deg cone around CubeSat-Sun direction).} Azimuth is the in-plane angle (Az=0~deg towards the projection of Sun's position on Didymos equatorial plane), elevation is the out-of-plane angle. Yellow and black markers indicate the direction of the Sun and Didymos, respectively, as seen from the release point. The blue marker represents the selected release direction.
           	\label{fig:release_azel}
        \end{figure}

        We performed an extensive Monte Carlo analysis of the parameter space (40k runs among eight difference release points) and, as a result, a release from 30~km (distance from Didymos system barycenter) in the direction (Az=10~deg, el=60~deg), is selected as a design baseline solution. Directions are shown in Didymos equatorial plane, where Azimuth is the in-plane angle (Az=0~deg towards the projection of Sun's position on Didymos equatorial plane), and elevation is the out-of-plane angle. Figure~\ref{fig:release_azel} shows the results of the analysis in terms of release direction, for the 30~km distant release point. Each release condition is propagated forward for 7~days. The compliance to the aforementioned constraints is checked at any time on the release trajectory. Feasible release directions are identified by green markers, while red markers indicate conditions where the constraints are not satisfied. Unfeasible release direction in Figure~~\ref{fig:release_azel} are characterized by a mix of green and red markers. This is due to effect of the release velocity: these directions are typically unfeasible for high release velocities (close to 5 cm/s), but might be feasible for lower velocities. In fact, the higher the release velocity, the higher the probability of falling beyond a 90~deg phase angle or 60~km distance from Hera after a 7-day ballistic arc. The blue marker highlight the selected release direction.

        \begin{figure}[t]
          	\centering
        	\includegraphics[width=\textwidth]{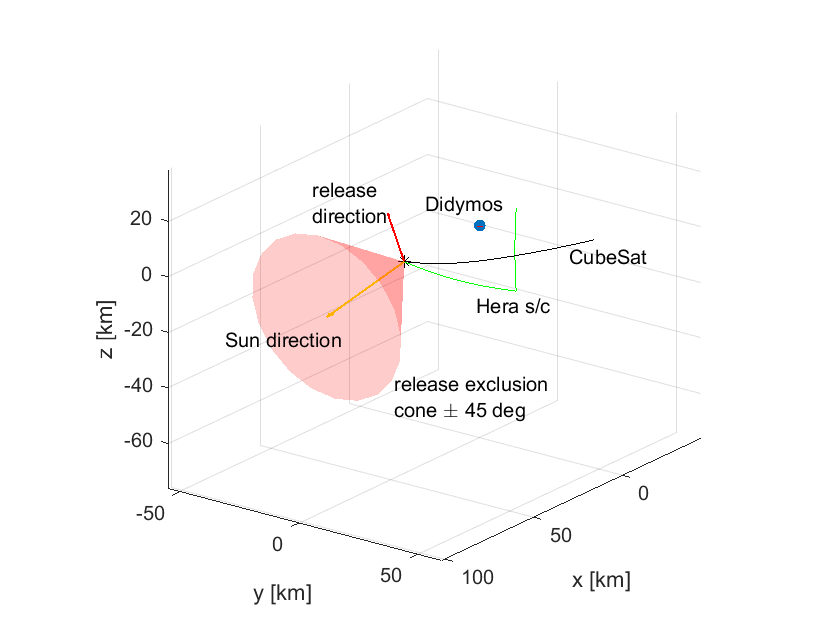}
           	\caption{Nominal release trajectory in the Didymos equatorial frame. Milani's and Hera's orbital motion is shown for seven days after release. Geometrical constraints (release exclusion angle around Sun's direction), and Didymos system are shown.}
           	\label{fig:release_traj}
        \end{figure}
        
        The release solution is shown in Figure~\ref{fig:release_traj}. The trajectory of Milani and Hera spacecraft are shown for seven days after release. This solution is compliant with design constraints, robust to uncertainties, and guarantees a safe 7-day ballistic arc after release.

    \subsection{Operational trajectories} 
    \label{sub:3.2-traj}
        After release from Hera and commissioning, Milani will maneuver to perform scientific operations. This section presents the selected operational orbit and maneuvering strategy to accomplish Milani's mission objectives. As summarized in Table~\ref{tab:Constraints}, scientific operations are carried at a distance within the range 4.572-10.940~km from Didymos system and when the phase angle (Sun-asteroid-CubeSat angle) of Didymos and Dimorphos is in the range 5-25~deg. Also, we enforce the CubeSat to have a non-zero elevation above Didymos equatorial plane, to ensure visibility of the asteroids poles. In terms of $\Delta v$ cost, we considered a budget of 10 m/s. In addition to constraints reported above, additional design drivers were considered, namely:
        \begin{itemize}
            \item \textbf{Safety}, in terms of orbital energy, quantified by the safety factor C. We give the highest priority to trajectories that are inherently safe, i.e. with C$>$0 (hyperbolic arcs). Closed trajectories (C$<$0), according to their range, have a moderate to high risk of entering chaotic three-body dynamics, potentially resulting in impacts with either of the bodies.
            \item \textbf{Simplicity}, in terms of reduced operational burden required to perform active operations on the CubeSat and optimization of the time allocated on Hera to relay Milani operations. To accomplish this, we leverage natural dynamics and perturbations of the asteroid environment to reduce the frequency of manoeuvres. In particular, we consider this criterium more important than $\Delta v$ cost, which is less critical in our mission.
            \item \textbf{Robustness} to uncertainties due to the system and dynamical environment. This is tightly connected to safety and simplicity criteria. Inherently safe trajectories are typically more robust to uncertainties. On the other hand, long ballistic time of flights between consecutive manoeuvres require robust trajectories, able to deal with uncertainties without jeopardizing the safety of the Hera mission or altering the mission profile strategy.
            \item \textbf{Cost}, in terms of $\Delta v$. Our analyses show that cost is not critical to our mission. For this reason, safe, simple, and robust trajectories are preferred, at a cost of a slightly higher $\Delta v$.
        \end{itemize}
        We performed an extensive trade-off analysis between several orbital strategies and trajectory solutions, including hyperbolic arcs, closed orbits and three-body proximity solutions. The results of the trade-off are summarized in Table~\ref{tab:tradeoff_matrix}. In particular, we show the suitability of each strategy in terms of available time for science operations (time in the science admissible region), navigation constraints (check on phase angles and asteroid illumination), safety/risk of entering chaotic dynamics, simplicity of operations/maneuvering strategy, dynamical robustness to uncertainties and $\Delta v$ cost. The study clearly identifies the hyperbolic-arc strategy as the most promising solution to host Milani scientific operations. This has several advantages in terms of safety, simplicity and robustness, and provides the best option in terms of asteroid visibility. In particular, we implemented a patched-arc manoeuvring strategy that leverage the SRP acceleration to target pre-selected waypoints. To reduce the burden for active operations, we consider a 4-3-4-3 maneuvering pattern, where maneuvers are performed after a 4- or 3-day ballistic arc. This is similar to the strategy implemented by the Hera spacecraft. Selecting such pattern has the advantage to ease operations at maneuvering points, which will be fit within a fixed weekly schedule and aligned to Hera operations.
        
        
        \begin{table}[t!]
            \centering
            \caption{Science orbit trade-off matrix.}
            \small
            \begin{tabular}{p{0.1\textwidth} | p{0.11\textwidth} p{0.12\textwidth} p{0.12\textwidth} p{0.12\textwidth} p{0.135\textwidth} p{0.09\textwidth}}
                Strategy & Time for science & Navigation & Safety/risk & Simplicity & Robustness & Cost: monthly\\
                \hline
                Loop (C$>$0)&
                \cellcolor{green!50} 1-2 days per loop&
                \cellcolor{green!50} Always on asteroids' dayside&
                \cellcolor{green!50} Inherently safe&
                \cellcolor{green!50} Compliant to 4-3-4-3 pattern&
                \cellcolor{green!50} 3-4 days&
                \cellcolor{orange!50} 1.5 m/s\\
                \hline
                Loop (C$<$0)&
                \cellcolor{green!50} 2-3 days per loop&
                \cellcolor{green!50} Always on dayside&
                \cellcolor{orange!50} Moderate risk&
                \cellcolor{green!50} Compliant to 4-3-4-3 pattern&
                \cellcolor{orange!50} 2-3 days&
                \cellcolor{orange!50} 1.2 m/s\\
                \hline
                Closed orbit (7~km)&
                \cellcolor{orange!50} $<$1 day per orbit&
                \cellcolor{orange!50} Dayside for half orbit&
                \cellcolor{orange!50} Moderate risk&
                \cellcolor{green!50} Compliant to 4-3-4-3 pattern&
                \cellcolor{orange!50} 2-3 days&
                \cellcolor{green!50} 20 cm/s\\
                \hline
                Closed orbit (2-7~km) &
                \cellcolor{red!50} $<$few hours per orbit&
                \cellcolor{orange!50} Dayside for half orbit&
                \cellcolor{red!50} High risk&
                \cellcolor{orange!50} Maneuver frequency 1-2 days&
                \cellcolor{red!50} $<$1 day&
                \cellcolor{green!50} $<$20 cm/s\\
                \hline
                3-Body Solution ($<$2~km)&
                \cellcolor{red!50} Never in science region&
                \cellcolor{red!50} Little dayside time; exceeding FOV&
                \cellcolor{red!50} High risk&
                \cellcolor{red!50} Maneuver frequency $<$1 day&
                \cellcolor{red!50} Not robust. Chaotic dynamics&
                \cellcolor{green!50} $<$20 cm/s\\
            \end{tabular}
            \label{tab:tradeoff_matrix}
        \end{table}
        
        \begin{figure}[t]
          	\centering
        	\includegraphics[width=\textwidth]{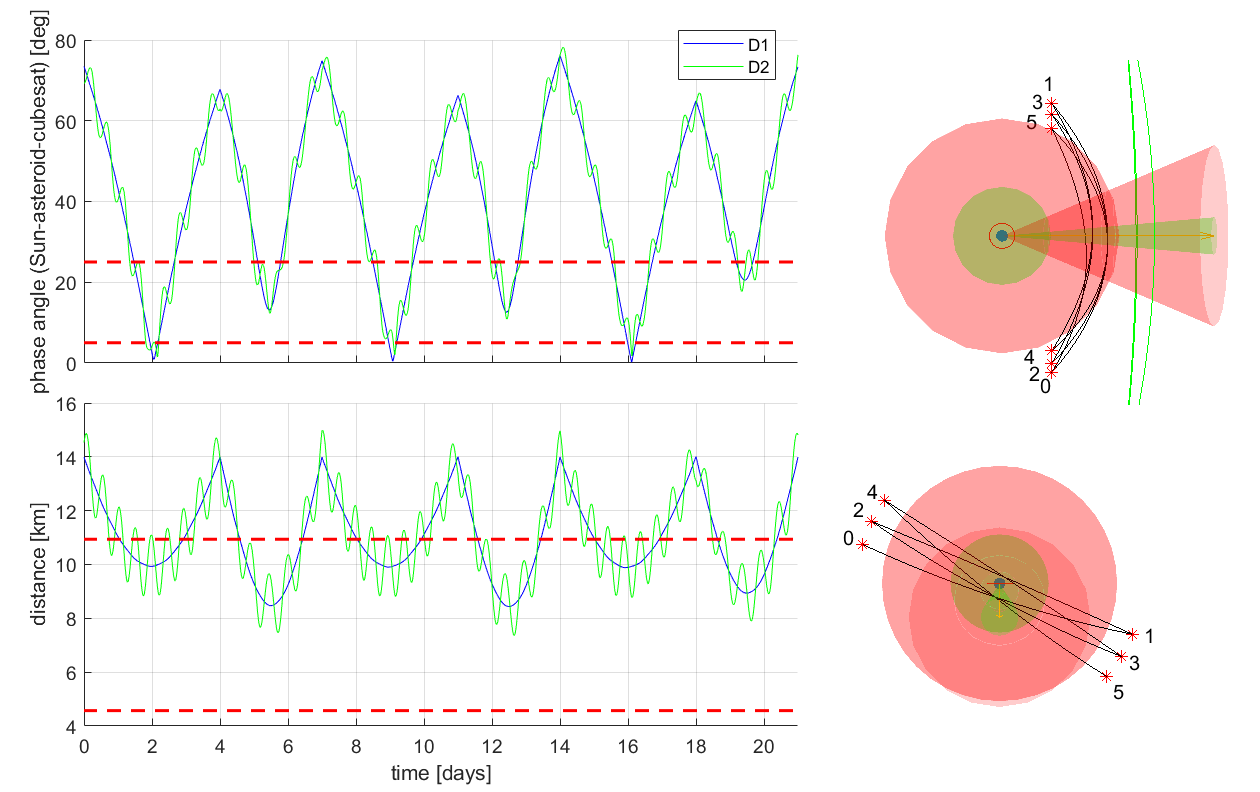}
           	\caption{Nominal science trajectory. Phase angle and distance time profile with respect to Didymos (D1) and Dimorphos (D2) are shown in upper and lower left plots. Science admissible region is highlighted (ranges between dashed lines). Projections on Didymos Equatorial frame x-y and y-z plane are shown in upper and lower right plots, respectively. These show science admissible regions in terms of distance from the binary system (green sphere is minimum distance, red sphere is maximum) and phase angles (green cone is minimum value, red cone is maximum value). The Sun direction is also shown (orange arrow inside the green cone), as well as the projection of Hera spacecraft's trajectory (green lines in the upper right plot).}
           	\label{fig:science_traj}
        \end{figure}    
            
        After a thorough design process, where several hyperbolic-arc waypoint configurations have been investigated, the waypoints of the hyperbolic loop are selected as shown in Figure~\ref{fig:science_traj}. Left plots in Figure~\ref{fig:science_traj} show the ranges and admissible science regions (between dashed lines). Milani has science windows of 1-2~days for each arc. Projections of the trajectory on the Didymos equatorial plane are shown on the right. To maximize the time spent by the spacecraft within the science admissible region, the orbital plane between consecutive hyperbolic arcs is tilted by a few degrees at each maneuvering point. This is clearly visible in the y-z projection plot (lower-right). This allows Milani to fly within admissible ranges in terms of aspect angle and distance, when transiting near the pericenter of each hyperbola. Numbered labels indicate the maneuver sequence, under the 4-3-4-3 day scheme, from maneuver 0 (insertion into science loop trajectory) up to maneuver 5. The waypoint design strategy is modular and can be further extended. As detailed in Section~\ref{sec:5-payload}, six hyperbolic arcs (21~days in total) are enough to accomplish the scientific objectives of the mission. If needed, the science loop trajectory can be further extended at the cost of 1.5~m/s per month.
        
        \subsubsection{End-of-Life options}
            Eventually, we discuss two End-of-Life (EoL) options in terms of their concept and feasibility in the broad context of the Hera mission, with the goal to identify benefits and criticalities. No baseline solution was selected within the context of this work.\\
            \underline{Option 1: Injection into a graveyard heliocentric orbit}. This option is safe and cheap, and it does not require any additional trajectory design after the scientific phase (except for a long-term integration of the outgoing heliocentric orbit). As discussed, the nominal science loop is built as a sequence of hyperbolic arcs. Missing one manoeuvre would safely bring Milani into an escape trajectory from the Didymos system. In addition, the SRP accelerates the CubeSat further away from Didymos, increasing its orbital energy. In this case, a safe graveyard heliocentric orbit can be achieved by missing the last manoeuvre of the scientific phase. After this, the spacecraft takes approximately 10~days to fly beyond a 60~km distance from Hera spacecraft. This option is the safest possible and may be preferred to avoid criticalities in terms of operational burden and costs. However, it does not provide any opportunistic science information in addition to those gathered during the science phase: it is a low risk-low gain solution.\\
            \underline{Option 2: Landing attempt on Dimorphos}. The second option is to attempt a landing on D2. After the nominal scientific and technological objectives of the mission are accomplished, it might be worth to exploit Milani to provide additional data on the Didymos system. To achieve this goal, a higher risk can be accepted. In this context, we plan to get closer to the Didymos system, dropping the design driver of flying inherently safe hyperbolic trajectories. As mentioned, orbiting the inner region of Didymos system poses several challenges in terms of navigation and robustness to manoeuvres. This is a high risk-high gain option. The knowledge analysis performed in Section~\ref{sub:4.2-knowl} shows that, in principle, the current GNC baseline would be compatible with a soft-landing design for the CubeSat. However, a more detailed analysis will be required to rigorously assess the feasibility of such option.

    \subsection{Momentum budget} 
    \label{sub:3.3-mombudget} 
        As mentioned, the SRP plays a fundamental role in assessing the dynamics of the CubeSat in the close-proximity of Didymos. Also, it contributes in a relevant manner to the active maneuvering costs to be provided to control the attitude of the CubeSat. The momentum budget is estimated here by assessing the impact of the SRP perturbation to the attitude dynamics of Milani. We use a cannonball model with an equivalent area of 0.5160~m$^2$, a reflection coefficient ($C_r$) of 1.1926, and a constant arm of 5~cm. These data are consistent for a 6U CubeSat with large solar arrays.
    
        \begin{figure}[t]
          	\centering
        	\includegraphics[width=0.9\textwidth]{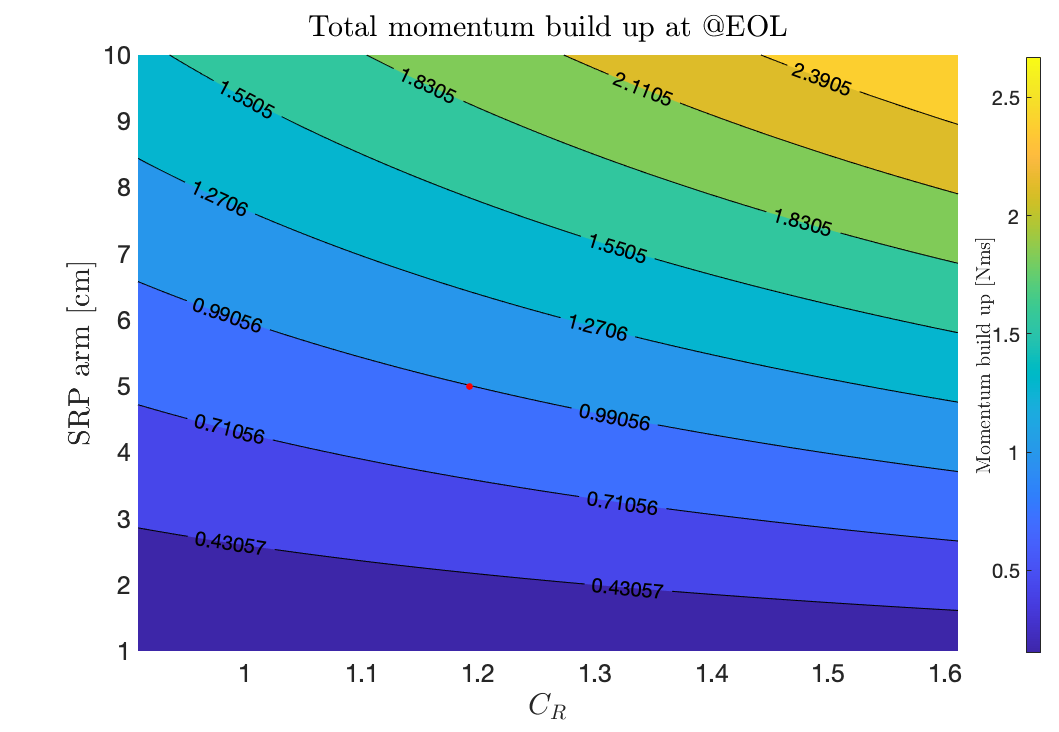}
           	\caption{SRP total momentum build up by the SRP on Milani after 6 months. The nominal value is shown with a red marker for $C_{R}=1.19$ and SRP arm=5~cm.}
           	\label{fig:SRP_momentum_buildup}
        \end{figure}
        
         As mentioned earlier, the baseline scenario considers that Milani is released in May 2027, when the distance from the Sun is 1.5~AU, reaching 1.9~AU and 2.2~AU after 3 and 6 months of operations, respectively. For comparison, we consider here the backup launch option as well, which considers a CubeSat release on January 2031, when the distance from the Sun is 1.1~AU, reaching 1.5~AU and 1.9~AU after 3 and 6 months of operations, respectively. The latter option is considered since it is a worst case scenario: the CubeSat is closer to the Sun and the effect of SRP is higher. The total momentum buildup due to SRP in the worst case scenario (backup launch and 6 months of operations) is illustrated in Figure~\ref{fig:SRP_momentum_buildup} as function of the key parameters of the cannonball model. With the nominal values a total momentum build up of 0.99~mNms during the whole mission is estimated.
    
\begin{figure}[t]
          	\centering
        	\includegraphics[width=0.9\textwidth]{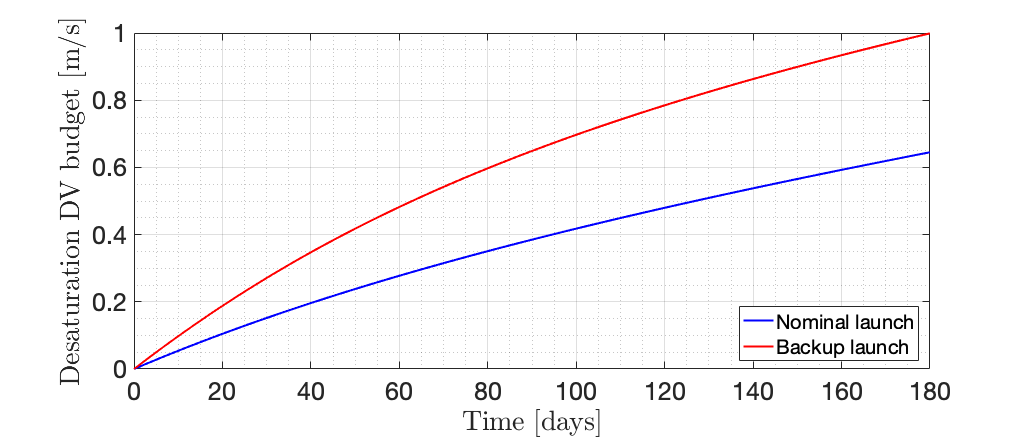}
           	\caption{Desaturation budget as function of mission duration for the nominal launch (blue curve) and backup launch (red curve).}
           	\label{fig:M_dump_DV_budget}
        \end{figure}

        Assuming that the disturbance torque is building up on a single reaction wheel, two options are considered in terms of momentum capacity: 25~mNms and 55~mNms. In the first case a total of 40 dumping manoeuvres would be required for 6 months of operations (approximately 1 manoeuvre every 4.5~days) while in the latter one the number drop to 18 for 6~months of operations (approximately 1 manoeuvre every 10~days). The total $\Delta V$ budget to be allocated for momentum dumping is therefore estimated to be around 1~m/s in the worst case scenario. The evolution of the $\Delta v$ budget for desaturation as function of the mission launch and duration is illustrated in Figure~\ref{fig:M_dump_DV_budget}.

\section{GNC strategy} 
\label{sec:4-gnc}
    The key driver for Milani's GNC subsystem design is to maximize the opportunities to observe the binary asteroid system. Also, it is paramount to complement the Orbit Determination (OD) with optical navigation, to always track the target asteroid, and to increase the efficiency of the operations by exploiting the pointing of the scientific observations for optical navigation. All in all, it is beneficial for the mission to never lose the line-of-sight to the target asteroid while performing scientific but also housekeeping tasks.  
    
\begin{figure}[t]
    \centering
         \includegraphics[width=1\textwidth]{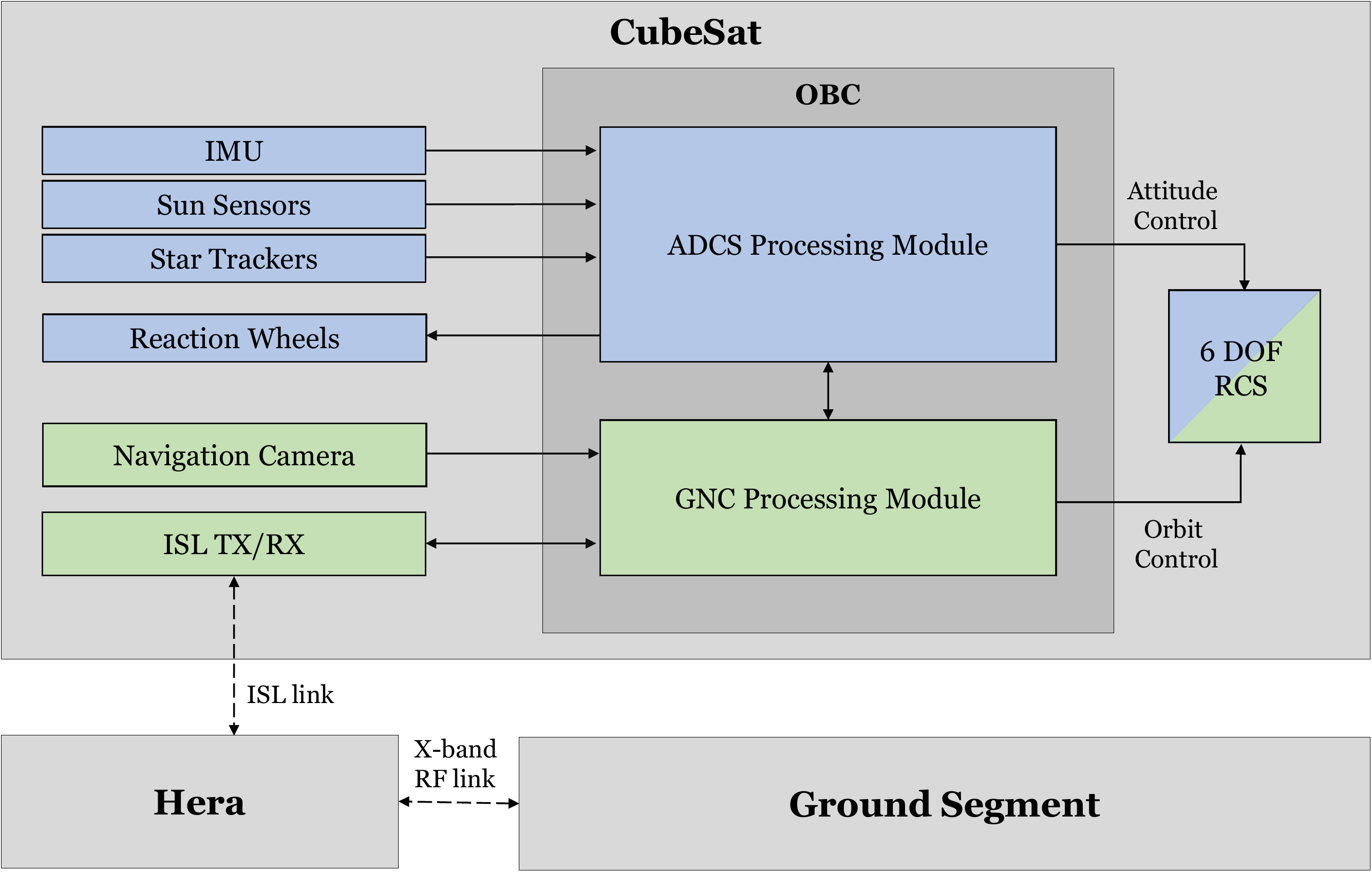}
         \caption{System architecture.}
         \label{fig:Architecture}
    \end{figure}

    The GNC system architecture is shown in Figure~\ref{fig:Architecture}. Milani is equipped with an on-board computer (OBC) where two modules are present, namely the ADCS processing module and the GNC processing module. These modules are responsible for collecting and processing sensor inputs to elaborate commands for the actuators. The  ground  segment  is  responsible  for computing the nominal guidance, navigation, and control of the satellite. This information is sent to the Hera spacecraft that is acting as a relay satellite and then sent back to Milani via the inter-satellite link (ISL). Apart from the ASPECT payload, we assume that Milani is equipped with a 21$\times$16 deg NavCam and a 40$\times$40 deg Wide Angle Camera (WAC). These are required for high-resolution imaging of the asteroids and optical navigation, as described in Section~\ref{subsec:opnav}.
    The navigation measurements for Milani come from the navigation camera and ISL. The navigation camera is used to acquire images of the two asteroids, while the ISL is exploited for range and range-rate measurements with respect to Hera. Milani is equipped with a Six Degree-of-Freedom Reaction Control System (6 DOF RCS) for trajectory and attitude control. The attitude measurements for Milani are given by an Inertial Measurement Unit (IMU), two sun sensors, and the star trackers. The attitude is also controlled by the reaction wheels.

    \subsection{Guidance} 
    \label{sub:4.1-guid}
        The Differential Guidance (DG) strategy  (\cite{dei2019lisa, park2006nonlinear}) adopted for Milani's guidance is detailed in this section. In the DG formulation, the whole trajectory is subdivided in different legs. At the extremal points of a single leg, two maneuvers are applied to cancel both the position and velocity deviations on the final leg point. However, the final impulse is usually not applied in practice, since at the time of arrival at the final point a new maneuver is calculated in a receding horizon approach. The maneuver can be computed by minimizing the deviations from the nominal state at the final point in a least square residual sense. Thus, the maneuver that has to be applied at the time $t_j$ in order to cancel out the deviations at time $t_{j+1}$ is computed as~\cite{dei2019lisa}
        \begin{equation}
        	\Delta\mathbf{v}_j=-\left(\Phi_{rv}^T\Phi_{rv}+\Phi_{vv}^T\Phi_{vv}\right)^{-1}\left(\Phi_{rv}^T\Phi_{rr}+\Phi_{vv}^T+\Phi_{vr}\right)\delta\mathbf{r}_j-\delta\mathbf{v}_j
        	\label{eq:DGDV}
        \end{equation}
        where $\delta\mathbf{r}_j$ and $\delta\mathbf{v}_j$ are the (estimated) position and velocity deviation at time $t_j$, $\Phi_{rr}$, $\Phi_{rv}$, $\Phi_{vr}$ and $\Phi_{vv}$ are the 3-by-3 blocks of the State Transition Matrix (STM) $\Phi\left(t_j,t_{j+1}\right)$ from time $t_j$ to time $t_{j+1}$ associated to the nominal trajectory. 
        
	    \subsection{Knowledge Analysis} 
        \label{sub:4.2-knowl}
        Errors in the dynamical propagation of the trajectories, e.g. inaccuracies in state determination or thrust misalignment, can lead to large deviations with respect to the nominal trajectory. Hence, a knowledge analysis is required to compute, through a covariance analysis, the achievable state knowledge. Simulations of the radiometric data for range and range-rate coming from the ISL are performed, generating the pseudo-measurements as
        \begin{equation}
            \rho = \sqrt{\boldsymbol{\rho}^T\boldsymbol{\rho}}+\varepsilon_\rho, \qquad \dot{\rho}=\frac{\boldsymbol{\rho}^T\boldsymbol{\eta}}{\rho}+\varepsilon_\eta
        \end{equation}
        where $\rho$ is the range, $\dot{\rho}$ is the range rate, $\boldsymbol{\rho}=\mathbf{r}-\mathbf{r}_H$ is the relative distance between Milani and Hera, while $\boldsymbol{\eta}=\mathbf{v}-\mathbf{v}_H$ is the relative velocity with $\varepsilon$ representing the error. $\mathbf{r}$ and $\mathbf{v}$ are position and velocity of Milani Cubesat, while $\mathbf{r}_H$ and $\mathbf{v}_H$ represent position and velocity of the Hera spacecraft.
        Pseudo-mesaurements are used to feed an Extended Kalman Filter~\citep[EKF,][]{schutz2004statistical} in order to simulate a realistic Orbit Determination procedure. In our case, the Hera spacecraft is always in view and no assessment of visibility windows is needed.
        
	\subsection{Guidance \& Navigation closed-loop approach}\label{sub:GNclosed}
        The overall navigation cost, necessary to keep the spacecraft on the nominal path, is estimated in a closed-loop fashion, taking into account:
        \begin{enumerate*}[label=\arabic*)]
            \item the knowledge analysis (as in Section \ref{sub:4.2-knowl}) to estimate the deviation of the real trajectory from the nominal trajectory at the target points;
            \item the differential guidance (as in Section \ref{sub:4.1-guid}) to assess the stochastic cost, starting from the estimated deviations.
        \end{enumerate*}
        The total navigation cost is estimated by means of a Monte Carlo simulation. Following this approach, an initial Gaussian distribution with mean ${\bar{x}}_0$ and covariance $P_0$ is identified and a set of samples $x_0^i$ is generated. Both the state and the covariance matrix are propagated with the associated dynamics up the first measurement epoch, where the estimated trajectory is updated. Proceeding in this way, the state estimates are sequentially updated as new measurements are processed, leading to the position and velocity knowledge profiles. This is repeated up to the final epoch of the Orbit Determination phase, where the deviation from the nominal path is estimated. The deviation is then pushed forward using the STM up to the maneuver time and used to feed the guidance law. The correction impulse is computed and applied. The whole process is repeated up to the final time for each initial state in the sample set. The estimation of the total cost for each sample is obtained as the sum of all maneuvers' cost, i.e., $\Delta v_i= \sum_{j=1}^{N} \left\|\Delta\mathbf{v}^i_j\right\|$.
    
    \subsubsection{Science phase}
        During the science phase, it is of paramount importance to acquire precisely and track the nominal trajectory in order to achieve the mission objectives.\\ 
        An assessment considering a higher fidelity model, with the OD process in the loop, is performed.
        Some assumptions are made for this analysis and they are listed below:
        \begin{itemize}
        	\item The guidance law is the standard differential guidance algorithm;
        	\item Correction maneuvers are given together with deterministic maneuvers;
        	\item No uncertainty in stochastic maneuvers is considered;
        	\item Due to technological constraints, correction maneuvers are not applied if their value is lower than 5~mm/s;
        	\item An initial a-priori uncertainty of 10~m on position, 1~mm/s on velocity (3$\sigma$) is used;
        	\item To compensate for differences between physical model and real world, a thrust misalignment of 1\% in magnitude and 1~deg in pointing angle (3$\sigma$) is considered;
        	\item The ISL is simulated taking into account an accuracy of 1~m in range and 0.1~mm/s in range-rate. The error is modelled as Gaussian noise;
        	\item In order to allow Flight Dynamics estimations and reduce the operational costs, an interval of 1.5~days between the maneuver and the beginning of the OD phase is considered, while a cut-off time of 1~day is inserted before the maneuver;
        	\item The ISL performs a range measurement every 20~minutes and a range-rate every 2~minutes;
        	\item Propagation is done linearly by means of the STM, while the estimation is done exploiting the EKF;
        	\item 1000 samples Monte Carlo run.
        \end{itemize}

\begin{figure}[t]
        	\centering
        	\includegraphics[width=1.2\textwidth]{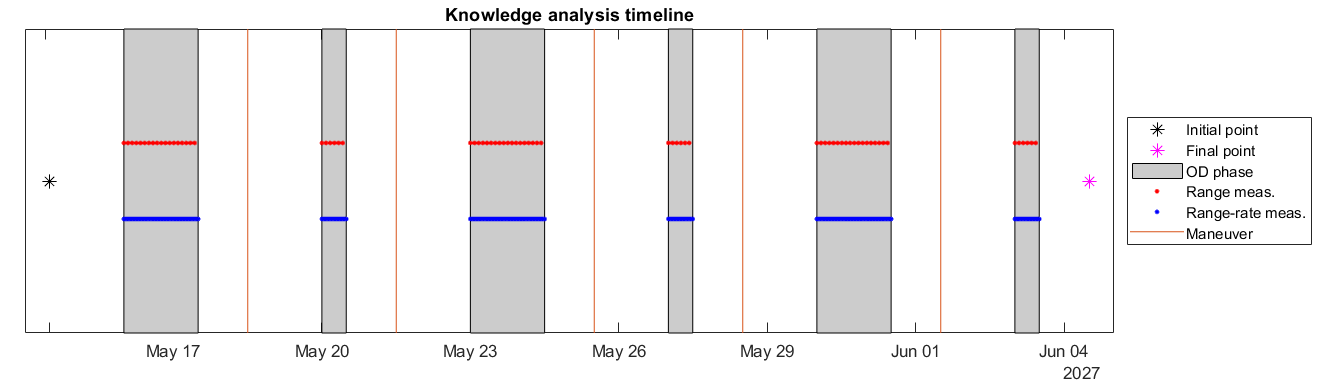}
        	\caption{Science orbit timeline (measurements dots are too close to be distinguished).}
        	\label{fig:timelineScience}
        \end{figure}

        Following these assumptions, a timeline for the science orbit can be derived. In Figure~\ref{fig:timelineScience}, red dots indicate when the range measurements are computed, while blue dots shows  range-rate measurements. Orange bars mark the maneuvers time.
    
\begin{figure}[t]
        	\centering
        	\includegraphics[width=0.8\textwidth]{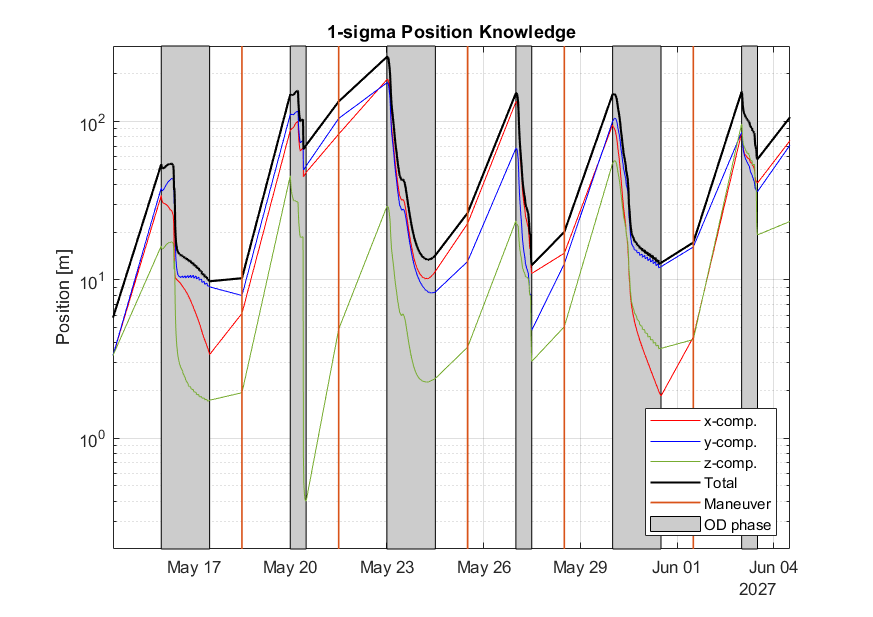}
        	\caption{Achievable position knowledge in the science phase.}
        	\label{fig:posScience}
        \end{figure}

\begin{figure}[ht]
    		\centering
    		\includegraphics[width=0.8\textwidth]{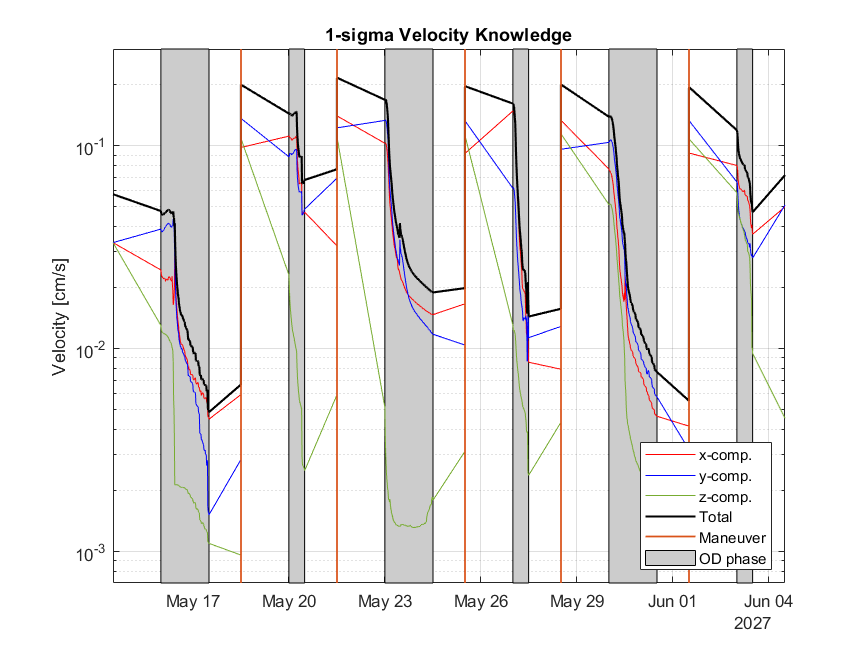}
    		\caption{Achievable velocity knowledge in the science phase.}
    		\label{fig:velScience}
    	\end{figure}

\begin{figure}[t]
        	\begin{subfigure}{.5\textwidth}
        		\centering
        		\includegraphics[width=1\linewidth]{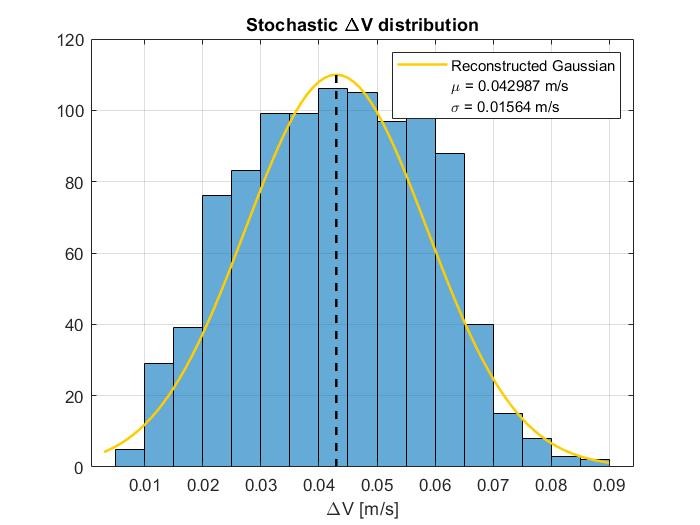}
        		\caption{}
        		\label{subfig:PDFScience}
        	\end{subfigure}
        	\begin{subfigure}{.5\textwidth}
        		\centering
        		\includegraphics[width=1\linewidth]{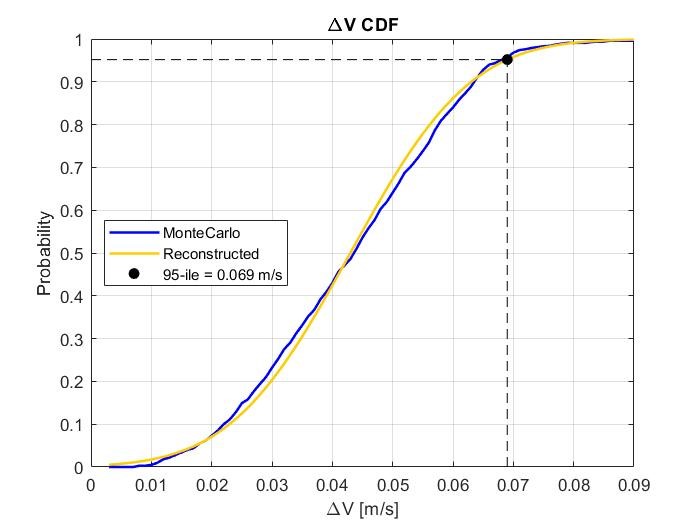}
        		\caption{}
        		\label{subfig:CDFScience}
        	\end{subfigure}
        	\caption{(a) Navigation cost probability distribution function for science phase. On the y-axis the number of occurrences are shown. (b) Navigation cost cumulative distribution function for science phase.}
        	\label{fig:DistScience}
        \end{figure}

\begin{figure}[t]
        	\begin{subfigure}{.5\textwidth}
        		\centering
        		\includegraphics[width=1\linewidth]{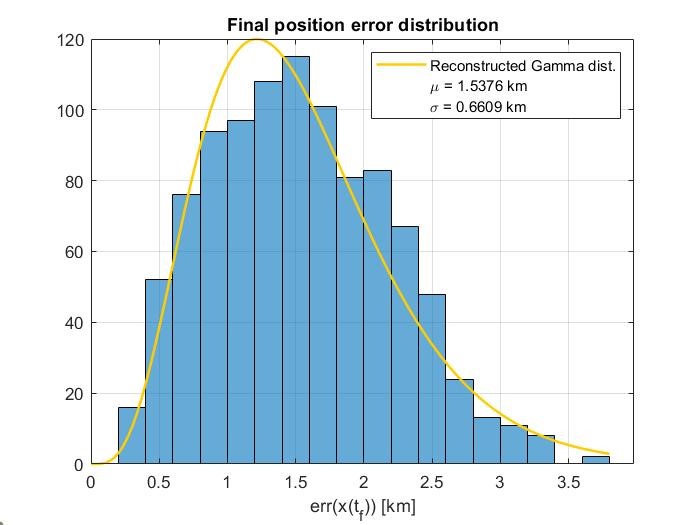}
        		\caption{}
        		\label{subfig:ErrPDFScience}
        	\end{subfigure}
        	\begin{subfigure}{.5\textwidth}
        		\centering
        		\includegraphics[width=1\linewidth]{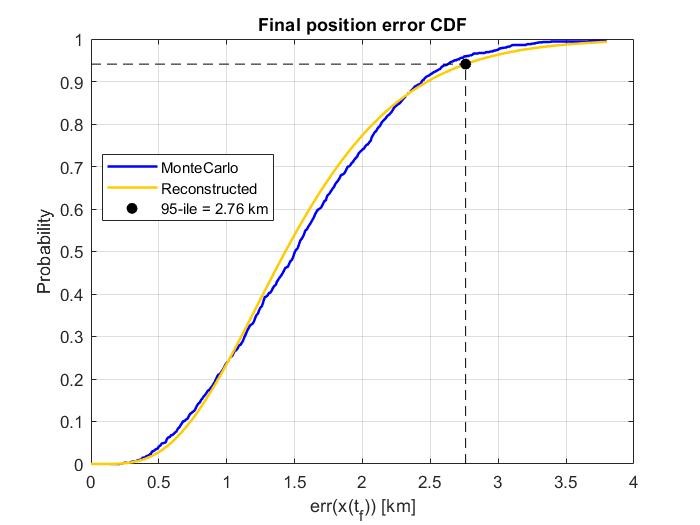}
        		\caption{}
        		\label{subfig:ErrCDFScience}
        	\end{subfigure}
        	\caption{(a) Final error probability distribution function for science phase. On the y-axis the number of occurrences are shown. (b) Final error cumulative distribution function for science phase .}
        	\label{fig:ErrDistScience}
        \end{figure}

        The outcome of the knowledge analysis is reported in Figure~\ref{fig:posScience} and Figure~\ref{fig:velScience}. They illustrate the results of the covariance analysis and, in particular, the achievable knowledge accuracy of the spacecraft during the science phase. In the plots, the DidymosECLIPJ2000 (centered in Didymos system barycenter, axes directed as ECLIPJ2000) reference frame is considered. At the manoeuvre points, a jump in the velocity covariance and a change of slope in the position covariance can be clearly noted. This effect is due to the uncertainty in the nominal maneuver. The high relative accuracy associated to the ISL leads a quite precise knowledge at the final point. Indeed, 1$\sigma$ position total accuracy is 100~m, while the velocity accuracy is better than 1~mm/s. Furthermore, the corrections are very effective in strongly reducing the dispersion. This leads to errors of some meters in position and few tens of microns per second in velocity after the correction step. The overall error associated to the OD is compatible with respect to the one assumed in the previous analyses when a mechanization of the error was considered. The 95\% confidence level for total correction maneuver cost in the science phase is 0.069~m/s (Figure~\ref{fig:DistScience}), while the position error one is 2.76~km, again mainly due to uncertainty in the final maneuver (Figure~\ref{fig:ErrDistScience}).
    
    \subsubsection{Release trajectory}\label{subsec:TP_Guidance}
        As for the release trajectory, the approach described in Section \ref{sub:GNclosed} is exploited. The first day after release is fully allocated to commissioning. After that, a first OD lasting 1~day for orbit acquisition is made. This is useful to reduce the huge dispersion given by the release. No correction maneuver is performed during the first 7~days. Then, a second Orbit Determination is performed from day 5 and day 6, where the deviation from the nominal path is estimated and then used as input for the differential guidance algorithm. In the second transfer phase leg, from the Injection Maneuver to the Science Orbit Acquisition Manuever (SOAM), the scheme similar to the science phase one is used: two correction maneuvers are given, the first after 4 days from the Injection Maneuver (IM) and the second after 7 days from the IM together with the SOAM. Differently from the science loop, the first correction maneuver is not associated with any deterministic maneuver, but it is needed in order to cope with the deviations caused by the Injection Maneuver and to target with an increased precision the Science Orbit Acquisition Maneuver point. In the second transfer phase leg a cut-off time of 1.5~days is considered between the OD and the maneuvers, plus a period of 1~day is taken into account between the maneuver and the beginning of the OD.
        For the initial state uncertainty, the worst-case scenario is taken as reference (see Section 3.3.1 for detailed discussion). Thus, a deployer release velocity of 5~cm/s with an uncertainty of 1~cm/s (3$\sigma$) is used. The half cone pointing error is 5~deg; on top of it the Hera pointing accuracy of 0.5~deg (3 $\sigma$) is considered.\\
        For clarity’s sake, the assumptions made for this analysis are summarized and listed below:
        \begin{itemize}
        	\item The guidance law is the standard differential guidance algorithm;
        	\item Correction maneuvers are given with the IM, the SOAM and at day 4 of the second leg;
        	\item No uncertainty in stochastic maneuvers is considered;
        	\item Correction maneuvers are not applied if their value is lower than 5~mm/s;
        	\item An initial uncertainty of 5~m on position is considered. The relative velocity with respect to Hera has a mean of 5~cm/s and a standard deviation of 1~cm/s (3$\sigma$). A pointing error of 5~deg (3$\sigma$, half cone) with a 0.5~deg pointing accuracy is used;
        	\item A thrust misalignment of 1\% in magnitude and 1~deg in angle (3$\sigma$) is considered;
        	\item The ISL is simulated taking into account an accuracy of 1~m in range and 0.1~mm/s in range-rate. The error is modeled as Gaussian noise;
        	\item During the first leg, Orbit Acquisition is performed from day 1 to day 2, while Orbit Determination from day 5 to day 6;
        	\item In the second leg, a interval of 1.5~days between the maneuver and the beginning of the OD phase is considered, while a cut-off time of 1~day is inserted before the maneuver;
        	\item The ISL performs a range measurement every 20~minutes and a range-rate every 2~minutes;
        	\item Propagation is done linearly by means of the STM, while the estimation is done exploiting the EKF;
        	\item 1000 samples Monte Carlo run.
        \end{itemize}

\begin{figure}[t]
        	\centering
        	\includegraphics[width=1.2\textwidth]{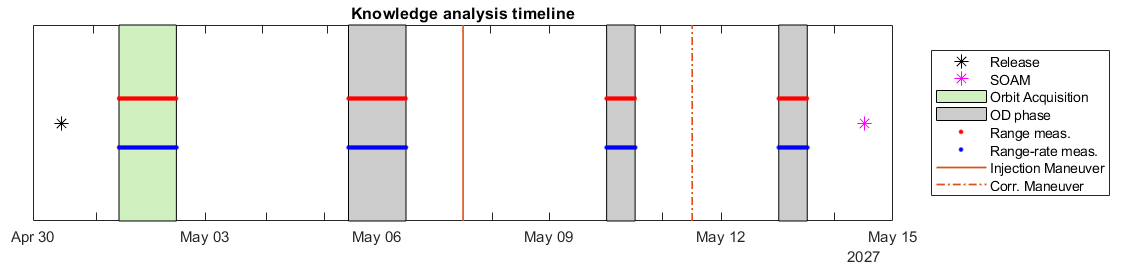}
        	\caption{Transfer trajectory timeline (measurements dots are too close to be distinguished).}
        	\label{fig:timelineTransfer}
        \end{figure}

        Following these assumptions, a timeline for the transfer orbit can be derived. In Figure~\ref{fig:timelineTransfer}, red dots indicates when the range measurements are computed, while blue dots shows when range-rate is performed.
    
\begin{figure}[t]
        	\centering
        	\includegraphics[width=0.8\textwidth]{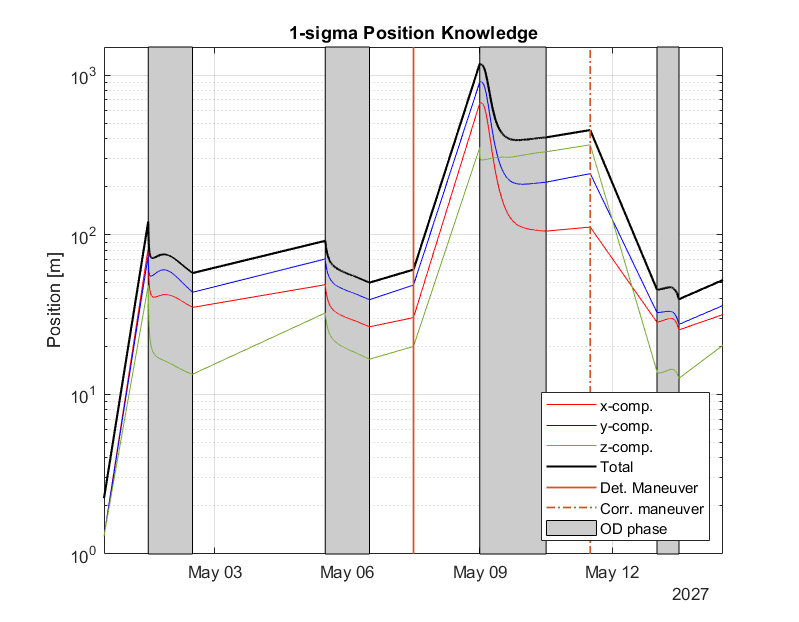}
        	\caption{Achievable position knowledge in the transfer phase}
        	\label{fig:posTrans}
        \end{figure}

\begin{figure}[t]
        	\centering
        	\includegraphics[width=0.8\textwidth]{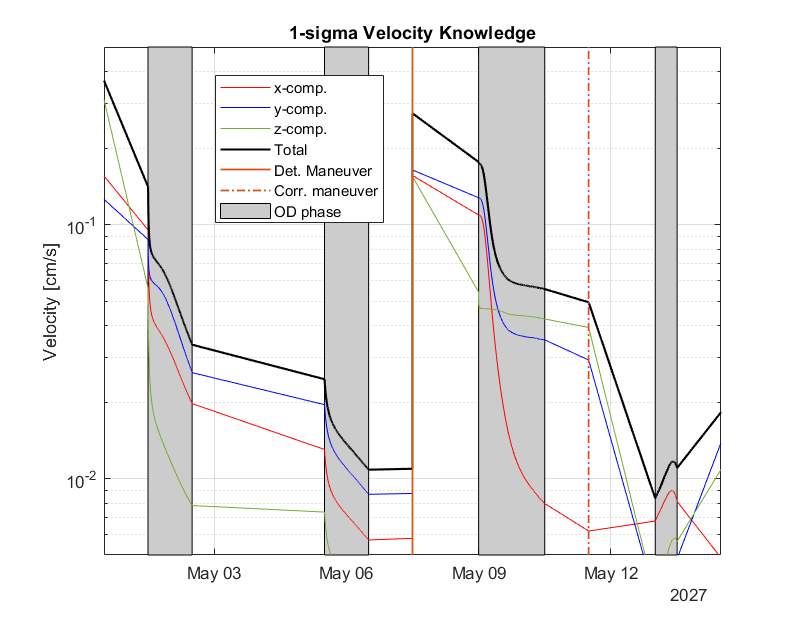}
        	\caption{Achievable velocity knowledge in the transfer phase}
        	\label{fig:velTrans}
        \end{figure}

\begin{figure}[t]
        	\begin{subfigure}{.5\textwidth}
        		\centering
        		\includegraphics[width=1\linewidth]{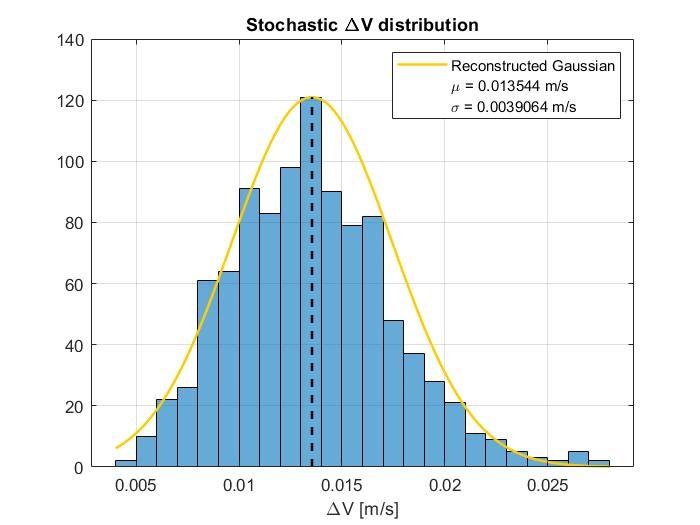}
        		\caption{}
        		\label{subfig:PDFTransfer}
        	\end{subfigure}
        	\begin{subfigure}{.5\textwidth}
        		\centering
        		\includegraphics[width=1\linewidth]{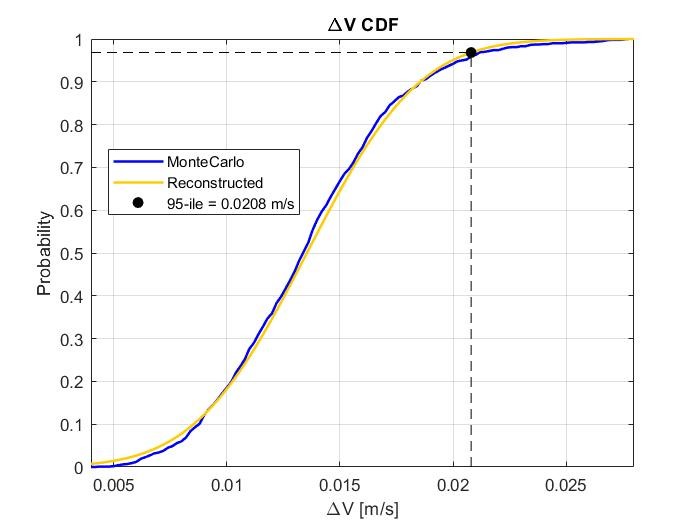}
        		\caption{}
        		\label{subfig:CDFTransfer}
        	\end{subfigure}
        	\caption{(a) Navigation cost probability distribution function for transfer phase. On the y-axis the number of occurrences are shown. (b) Navigation cost cumulative distribution function for transfer phase.}
        	\label{fig:DistTransfer}
        \end{figure}

        The knowledge analysis is reported in Figure~\ref{fig:posTrans} and Figure~\ref{fig:velTrans}. They illustrate the results of the covariance analysis and the achievable knowledge accuracy of the spacecraft during the transfer phase. In the plots, the DidymosECLIPJ2000 reference frame is considered. The large dispersion at the beginning is only partially recovered through the Orbit Determination and an error of some tens of meters is found for the position at the SOAM. On the other hand, the velocity knowledge improves better during the transfer, leading to error less than 0.1~mm/s at the end of transfer phase. These figures show how the orbit acquisition phase is beneficial to enhance the knowledge analysis in the first transfer leg, quenching the dispersion and helping in determine the correct orbit the spacecraft is on just after the release. The covariance analysis shows also that the navigation costs (Figure~\ref{fig:DistTransfer}) can be represented as a Gaussian distribution of mean 0.01354~m/s and a standard deviation of 0.0039~m/s. The 95\%-ile for correction maneuver cost in the transfer phase is 2.1~cm/s.
    
    \subsection{Optical Navigaton} \label{subsec:opnav}
        The baseline navigation strategy exploits information of range and range-rate with respect to the mother spacecraft. In addition, the mission will perform an optical navigation experiment by processing the images of the asteroids. This is intended to validate the performances of optical navigation on-board, but will not impact the baseline navigation strategy.
        
        \subsubsection{State-of-art methods and trade-off} 
\begin{figure}[t]
        \centering
             \includegraphics[width=0.4\textwidth]{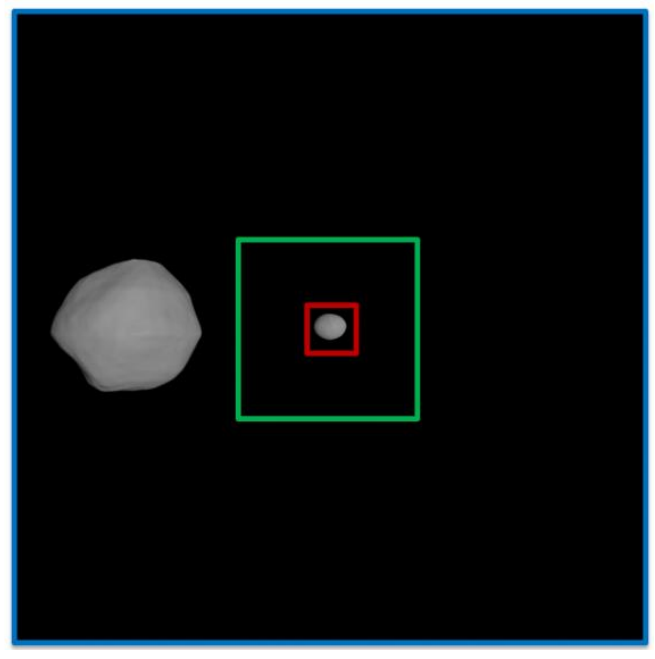}
             \caption{The external frame represents the FOV of the WAC, the medium frame represents the FOV of the NavCam, while the small asteroid is included in the small ASPECT field-of-view. The distance at which this image is simulated is about 8.5 km from Didymos.}
             \label{fig:FOVS}
        \end{figure}

        Different state-of-art strategies can be adopted for optical navigation. They range from the landmark-based methods to centroiding and disk fitting methods. The landmark navigation estimates the position of Milani based on the natural landmarks on the asteroid surface \cite{PardoRosetta}. It is accurate, however is also heavy from a computational standpoint. The centroiding method with disk fitting exploits the acquisition of the center of brightness of the figure to estimate the relative line-of-sight and the external horizon to estimate the apparent asteroid size, which is related to the relative range \cite{Gil-FernandezAutonomous}. This method can yield a good accuracy, the image processing required is simple, and an accurate asteroid model is not required. The horizon-based navigation is similar to landmark navigation, but it compares the asteroid shape and dimension as seen in the images with an asteroid model \cite{OwenMethods}. The correlation between the two yield an information about the relative position and pose of Milani at a very high computational cost, and an accurate asteroid shape model is required.  We performed a trade-off analysis to select the most suitable method and the centroiding with disk fitting resulted as the most promising one for optical navigation owing to its accuracy, robustness, simplicity, and independency from the asteroid model knowledge. Thus, a navigation strategy exploiting the centroiding technique with disk fitting is baselined for navigation. During the occultation periods of the secondary by the primary body or shadow, the navigation strategy will still be robust, as Milani could either use ISL range and range-rate measurements or use Didymos visual images (provided that the navigation NavCam field-of-view will be able to image Didymos fully within the frame) or relying on on-board propagation to avoid losing the tracking of the system. However, since the shape of the asteroids and their features are not known, the centroiding method with a disk fitting has been chosen as the baseline for optical navigation. In order to assure accuracy, observability of the whole system, coverage, and efficiency of operations, the navigation strategy uses a different navigation camera which has a larger field-of-view and higher resolution than ASPECT. In this way, while performing scientific operations, it is still possible to acquire navigation information. Moreover, a separated wide-angle camera (WAC) is foreseen to grant the full system visibility at any time. Thus, the adopted strategy for optical navigation exploits the same target of the scientific investigation with a navigation camera but keeps the observability of the two asteroids with a small ultra-wide angle camera. In this way, the pointing is shared among science and navigation, the navigation observables are accurate because the navigation camera is dedicated to the asteroid, and the full binary system view is assured owing to the ultra-wide angle camera. This strategy is also beneficial during the safe mode and targets acquisition mode. The parameters of the navigation cameras are reported in Table~\ref{tab:NavCams}. Due to the early stage of the design process, the GNC baseline has not been established yet in terms of hardware components. The values in Table~\ref{tab:NavCams} are assumed based on typical performance of similar sensors, and do not refer to any existing camera. A representative image of the Didymos system with the three different cameras (ASPECT, NavCam, and WAC) is shown in Figure~\ref{fig:FOVS}.

        \begin{table}[t]
            \centering
            \caption{Navigation camera characteristics.}
            \begin{tabular}{lcc}
                \toprule
                \toprule
                \textbf{Camera} & \textbf{FOV} & \textbf{Sensor} \\
                \midrule
                NavCam & 21 x 16 deg & 2048 x 1536 pix \\
                WAC & 40 x 40 deg & 2048 x 2048 pix \\
                \bottomrule
            \end{tabular}
            \label{tab:NavCams}
        \end{table}

        \subsubsection{Optical Navigation Model and Methods}
\begin{figure}[t]
            \centering
                \includegraphics[width=1\textwidth]{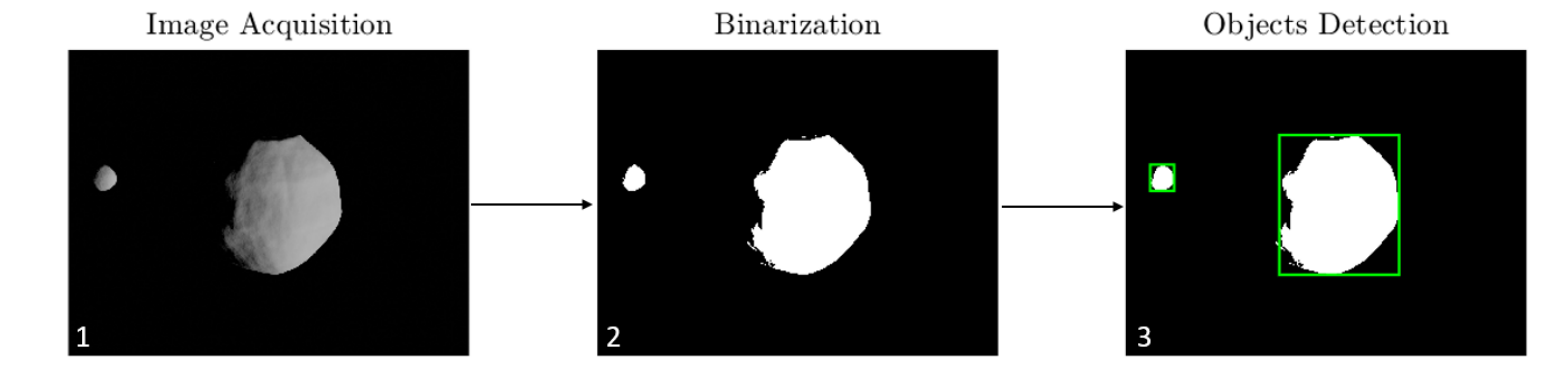}
                \caption{Steps 1-3; Image acquisition, pre-processing, and objects detection.}
                \label{fig:ImgProc1}
            \end{figure}

            The image processing steps and the optical navigation method are described in this paragraph. The image, generated according to the CubeSat trajectory, pointing, and navcam specifications, is retrieved by Celestial Objects Rendering Tool (CORTO), which is a tool developed at Politecnico di Milano (Figure~\ref{fig:ImgProc1}, step~1). The image is pre-processed to compute a mean background noise which is subtracted from the image, and the image is thresholded to highlight the bright objects in the frame (Figure~\ref{fig:ImgProc1}, step~2). In this way, it is straightforward to identify the two groups of connected pixels belonging to D1 and D2, which can be bounded by a green box (Figure~\ref{fig:ImgProc1}, step~3).
            
\begin{figure}[t]
            \centering
                \includegraphics[width=1\textwidth]{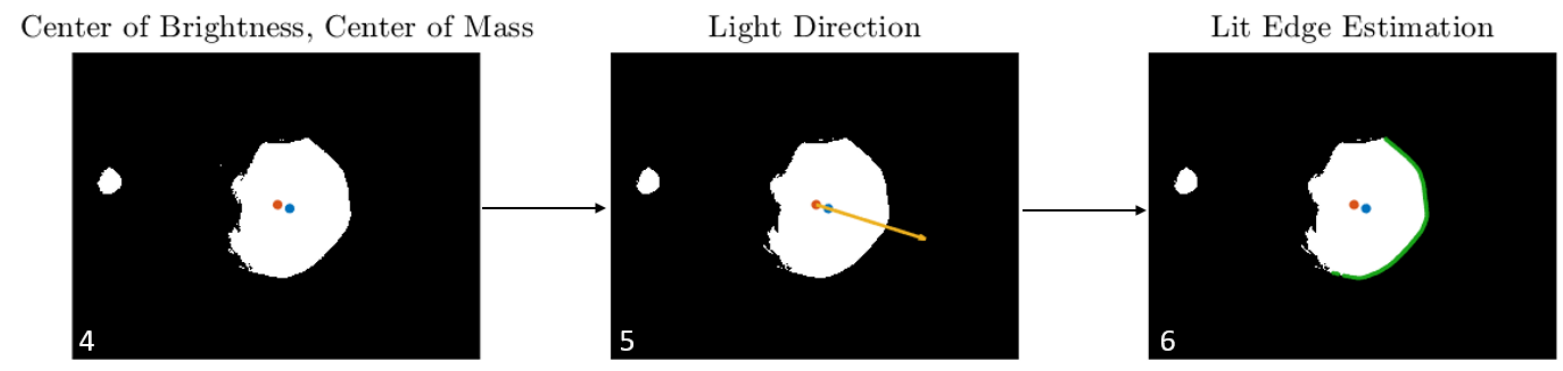}
                \caption{Steps 4-6; Center of brightness (blue dot), center of mass (red dot), light direction and lit edge estimations.}
                \label{fig:ImgProc2}
            \end{figure}

            The center of brightness of each object can be easily computed by determining the centroid of the illuminated pixels inside the green boxes (Figure~\ref{fig:ImgProc2}, step~4). The principal axes of each object can be determined, one of this axis will be aligned with the light direction while the other in the direction orthogonal to it. In this way, the axis of the light direction can be estimated from the images but it can be also complemented by the sun sensor information (Figure~\ref{fig:ImgProc2}, step~5). The center of mass can be then estimated as a displacement of the center of brightness along the negative light direction. The displacement between the center of brightness and the center of mass is inversely proportional to the object phase angle (Figure~\ref{fig:ImgProc2}, step~5). The image is then scanned along the light direction to identify the pixels belonging to the lit edge of each body (Figure~\ref{fig:ImgProc2}, step~6). 

\begin{figure}[t]
            \centering
                \includegraphics[width=1\textwidth]{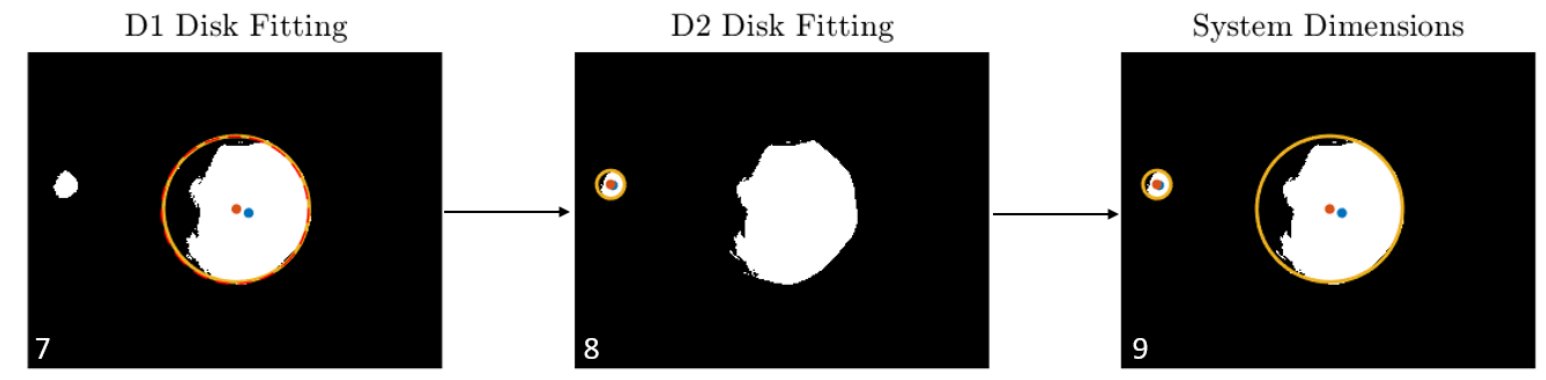}
                \caption{Steps 7-9; D1 disk fitting and D2 disks fitting.}
                \label{fig:ImgProc3}
            \end{figure}

            Owing to the estimations of both the center of mass and the lit edge of each object in the images, a mean radius in pixels can be estimated from the center of mass to the pixels constituting the lit edge. In this way a disk is fitted to the object (Figure~\ref{fig:ImgProc3}, step~7 for D1 and step~8 for D2). The center of mass, the center of brightness, and the dimensions of the two objects are the outputs that can be extracted from the image processing algorithm.

        \subsubsection{Image Processing Accuracy}
\begin{figure}[t]
            \centering
                 \includegraphics[width=1\textwidth]{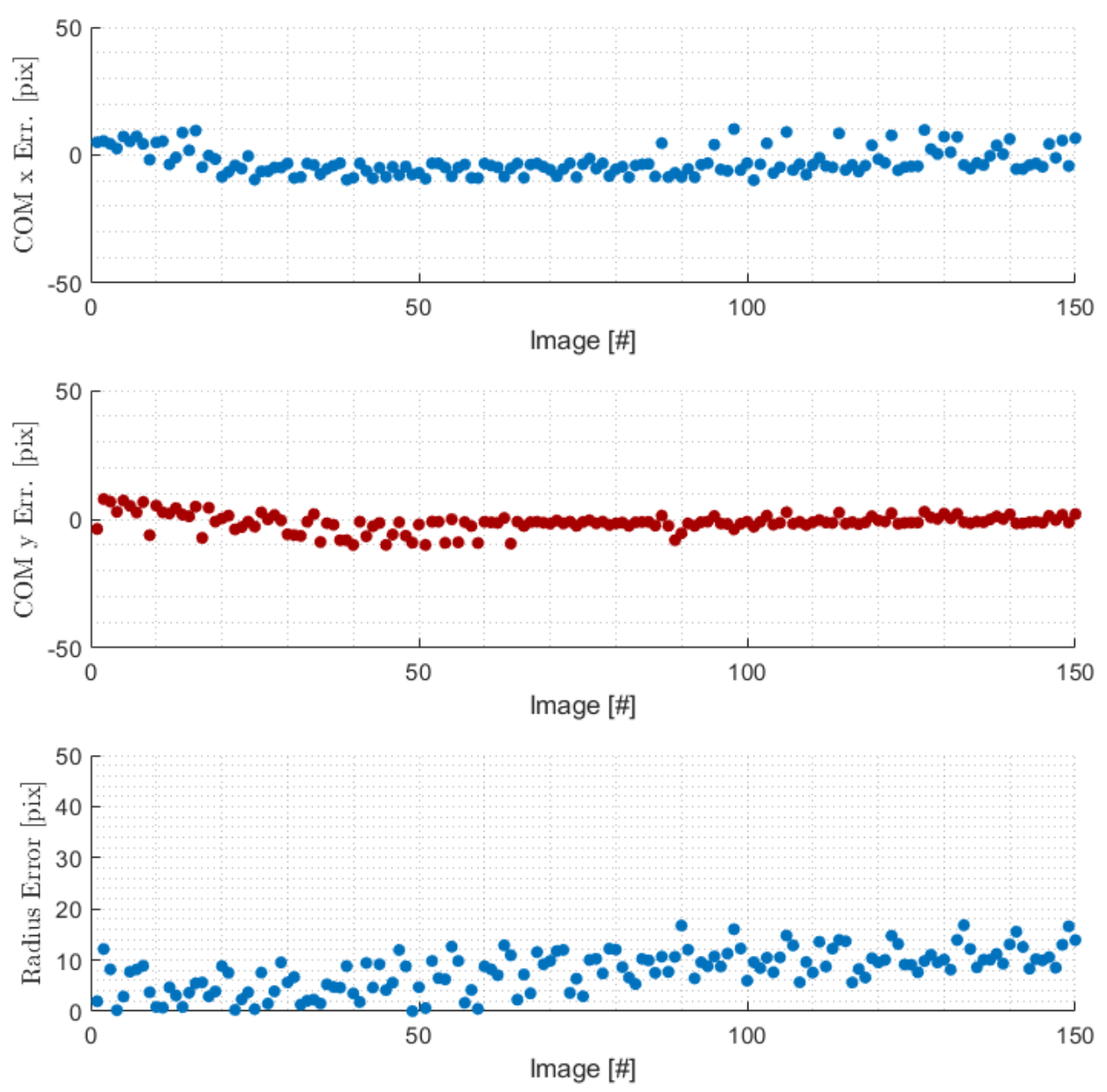}
                 \caption{Center of mass error in horizontal and vertical pixel directions $x$ and $y$ and radius error as output of the image processing.}
                 \label{fig:NavOutputs}
            \end{figure}

            The optical navigation method accuracy is evaluated by processing 150 images randomly chosen from the nominal orbit. The procedure extracts navigation information from the figures in a static way, thus no dynamic filtering is used in this preliminary phase. The image processing steps extract the information on the centroid of D1 and the apparent disk size. This information are given in pixels and are compared to the actual ones to assess the errors involved with the image processing. Figure~\ref{fig:NavOutputs} shows the mentioned errors, where the $x$ and $y$ directions are the horizontal and vertical pixel coordinates and COM refers to the center of mass. The error in the centroiding has a 3$\sigma$ std lower than 10~pixels, while the error in the disk radius estimation has a 3$\sigma$ std lower than 20~pixels. 
 
            The NavCam has a 21$\times$16~deg FOV and a 2048$\times$1536 pixels detector. Thus, since every pixel spans 0.0103~deg, the worst case centroiding error is 0.103~deg. This is equivalent to a lateral error of 12.5~m at a distance of 7~km, and a lateral error of 17.9~m at a distance of 10~km.
            The worst case pixel radius error is 20 pixels (equivalent to a 0.2051~deg error). This is equivalent to a range error of 423.8~m at true distance of 7~km, and to a range error of 841.9~m at a true distance of 10~km. 
            Thus, the worst case range error coming from the image processing is lower than the 10$\%$ of the true range. The error might be still reduced by developing an on-board optical navigation filter.
 
    \subsection{Control} 
    \label{sub:4.3-contr}
        Milani's nominal control architecture is expected to be based on traditional methods. However, because of the unique operations of Milani, in concert with other spacecraft (Hera and Juventas) and the risk of collision with either of the asteroids, particular care is involved in the design of the Corrective Action Maneuver (CAM) strategy. The CAM is a semi-autonomous contingency maneuver which is executable during nominal operations, in this section the strategy involved in the evaluation, quantification and execution of the maneuver is explained.
        
        \begin{figure}[t]
          	\centering
        	\includegraphics[width=0.8\textwidth]{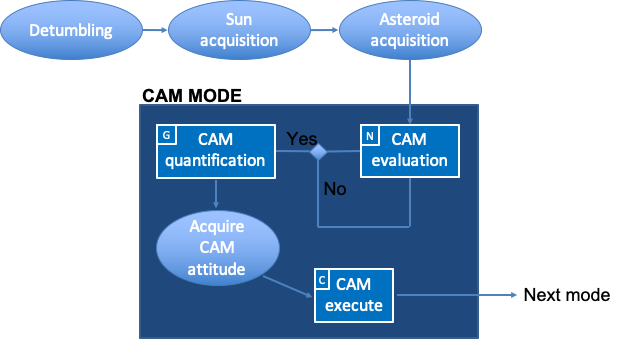}
           	\caption{High-level architecture of the CAM strategy.}
           	\label{fig:CAM_arch}
        \end{figure}
    
        During operations Milani will be sharing the most valuable space around Didymos system with Juventas and Hera. To ensure safe operations, the CubeSat shall be capable to avoid collisions with the surrounding bodies. The safety of Milani starts at mission analysis level with the design of trajectories that nominally do not encroach with the spacecraft and asteroids. However, a high-level strategy for the CAM is designed for unsafe deviations from nominal operations. Its schematics is shown in Figure~\ref{fig:CAM_arch}. It is desirable to avoid both collisions with the spacecraft and with the asteroids. However, it is also important to distinguish between these two different events. The happening of the first one would produce the loss of multiple systems, while the latter would involve Milani. The collision risk can be assessed directly with the ISL range and range-rate capabilities for the spacecraft (Hera and Juventas, both active systems) while a combination of visual images from the NavCam and ISL data can be used with the asteroids (Didymos and Dimorphos, both passive systems).
    
\begin{figure}[t]
          	\centering
        	\includegraphics[width=0.9\textwidth]{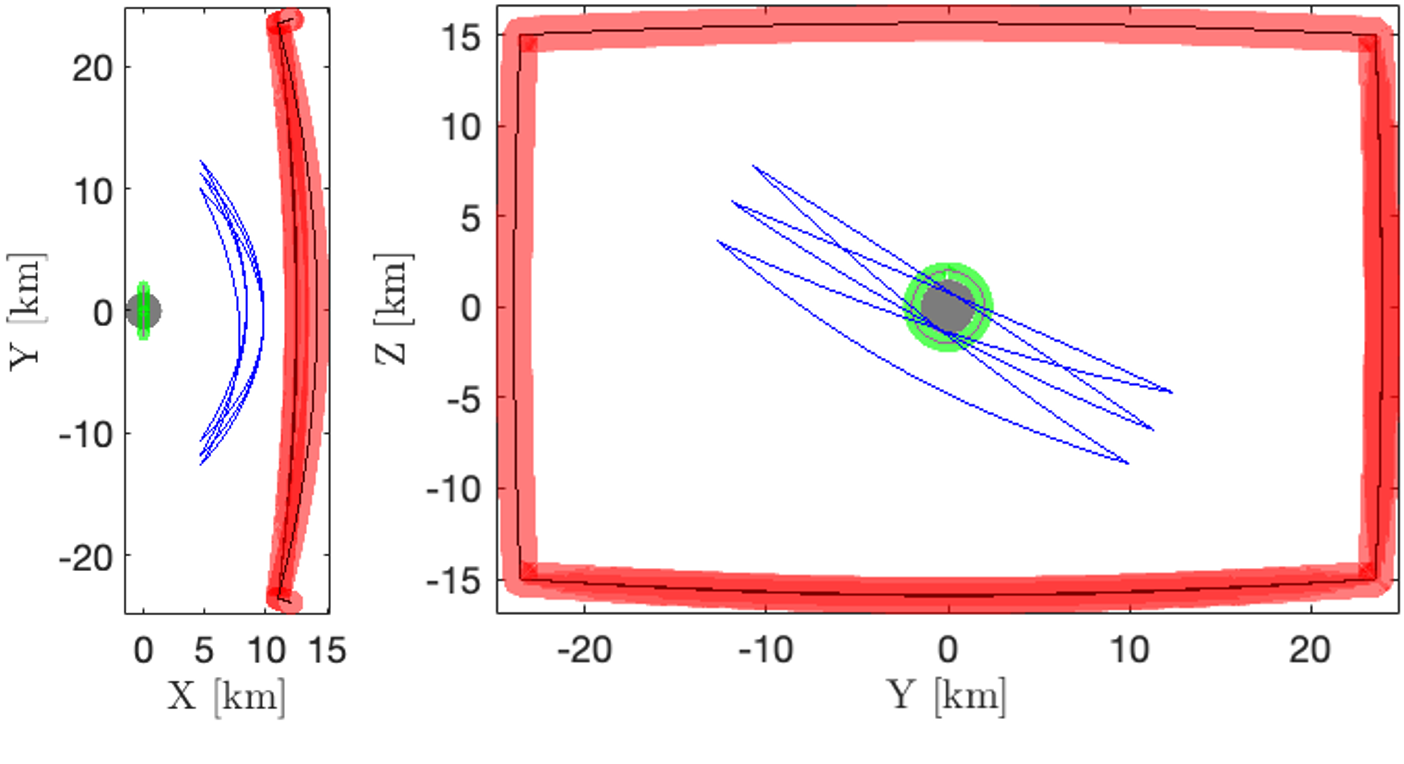}
           	\caption{Example of risk tubes during the science phase of Milani mission. The trajectories of Hera (black rectangle), Juventas (inner magenta circle) and Milani (blue) are represented in the Didymos Equatorial reference frame. The risk tubes for Hera and Juventas are represented with radius respectively of 1~km and 500~m. A contingency spherical region of risk is assumed for Didymos with a radius of 1.5~km.}
           	\label{fig:CAM_tubes}
        \end{figure}

        The triggering mechanism of the CAM mode is based on the concept of risk ranges, that are simply the estimated ranges from the spacecraft and the asteroids. Whenever these quantities go below a certain threshold, a correction maneuver is considered necessary. Figure~\ref{fig:CAM_tubes} shows the risk ranges associated to Hera, Juventas and the Didymos system. Their evolution in time is describing risk tubes.
    
        Once a corrective maneuver is requested, its orientation and magnitude are computed. The strategy to do that on-board is based on an hybrid between a generic and a specific look-up tables. The generic one contains only the magnitude of the maneuvers as a function of the range of Milani from the colliding object. The delta-V vector of the maneuver is directed towards the local radial direction of Milani's osculating orbit, opposite to the asteroid system. These maneuvers are pre-computed on ground based on the dynamic environment of the asteroid in such a way to ensure Milani will get further away but will not escape the system before a certain amount of days to be specified. The specific one contains pre-computed maneuvers both in magnitude and direction that are evaluated with a certain timespan on the nominal trajectory. The specific one are more accurate and can be relevant for small deviations from the nominal trajectory, the generic one are less accurate but are designed to cover the entire space around the colliding objects.

\section{Payload operations} 
\label{sec:5-payload}
    In this section a detailed analysis of payload operations during the scientific phase of the mission is illustrated. In particular, we only focus on ASPECT operations, which is the only payload that impose observational requirements to the mission profile. The driving requirements are in terms of phase angle and resolution. Together with resulting distances, these are summarized in Table~\ref{tab:DidymosSystem} and Table~\ref{tab:Constraints}.
    

    
    A science orbit cycle of 21 days is considered in the analysis to assess the capability to fulfill the payload objective to produce global maps of the asteroids. The time intervals within this cycle in which the constraints on the phase angle, distance and resolution are simultaneously satisfied are considered for potential observations.
     
    Further pruning is performed by assessing the illumination condition and viewing geometry of each individual face of the asteroid shape models. Shadowing effect of Didymos on Dimorphos are also taken into account. Considering all these factors, it is estimated that a total cumulative time of 4.7 and 4.6~days are available to perform scientific observations of, respectively, Didymos and Dimorphos. There are however regions in both bodies near the south poles that will be permanently in shadow during the mission. These cannot be observed by ASPECT using visual imaging and are not caused by the choice of Milani trajectories, and therefore are not accounted in the evaluation of the global coverage. 

    \begin{figure}[p]
      	\centering
    	\includegraphics[width=0.92\textwidth]{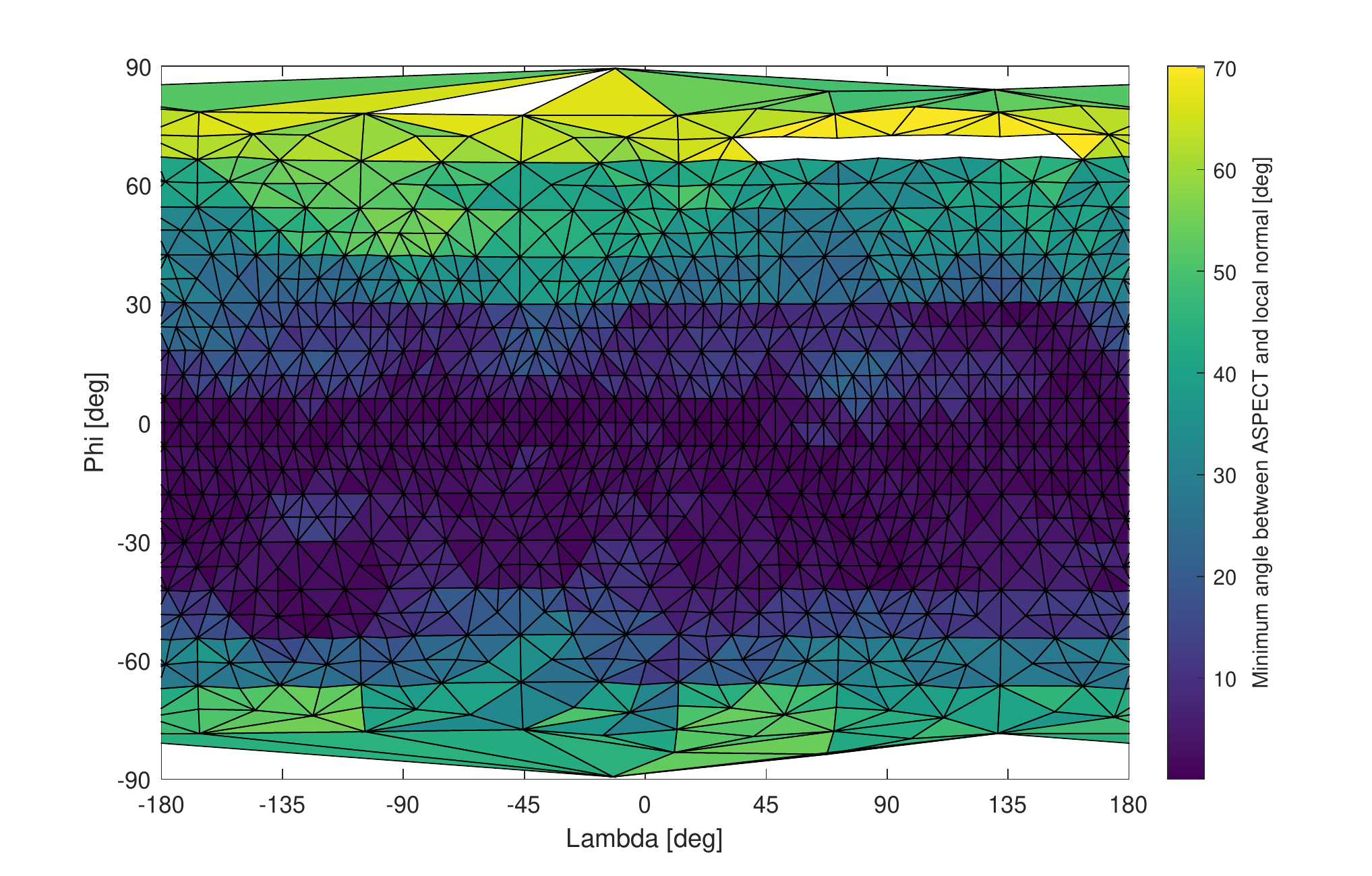}
       	\caption{ASPECT potential global coverage on Didymos. The map represents which faces of the shape model can be imaged by the payload given that the observation requirements are satisfied. The color of each face represents the minimum angle between the ASPECT line of sight and the face local normal during the science orbit cycle.}
       	\label{fig:coverage_D1}
    \end{figure}
    
    \begin{figure}[p]
      	\centering
    	\includegraphics[width=0.92\textwidth]{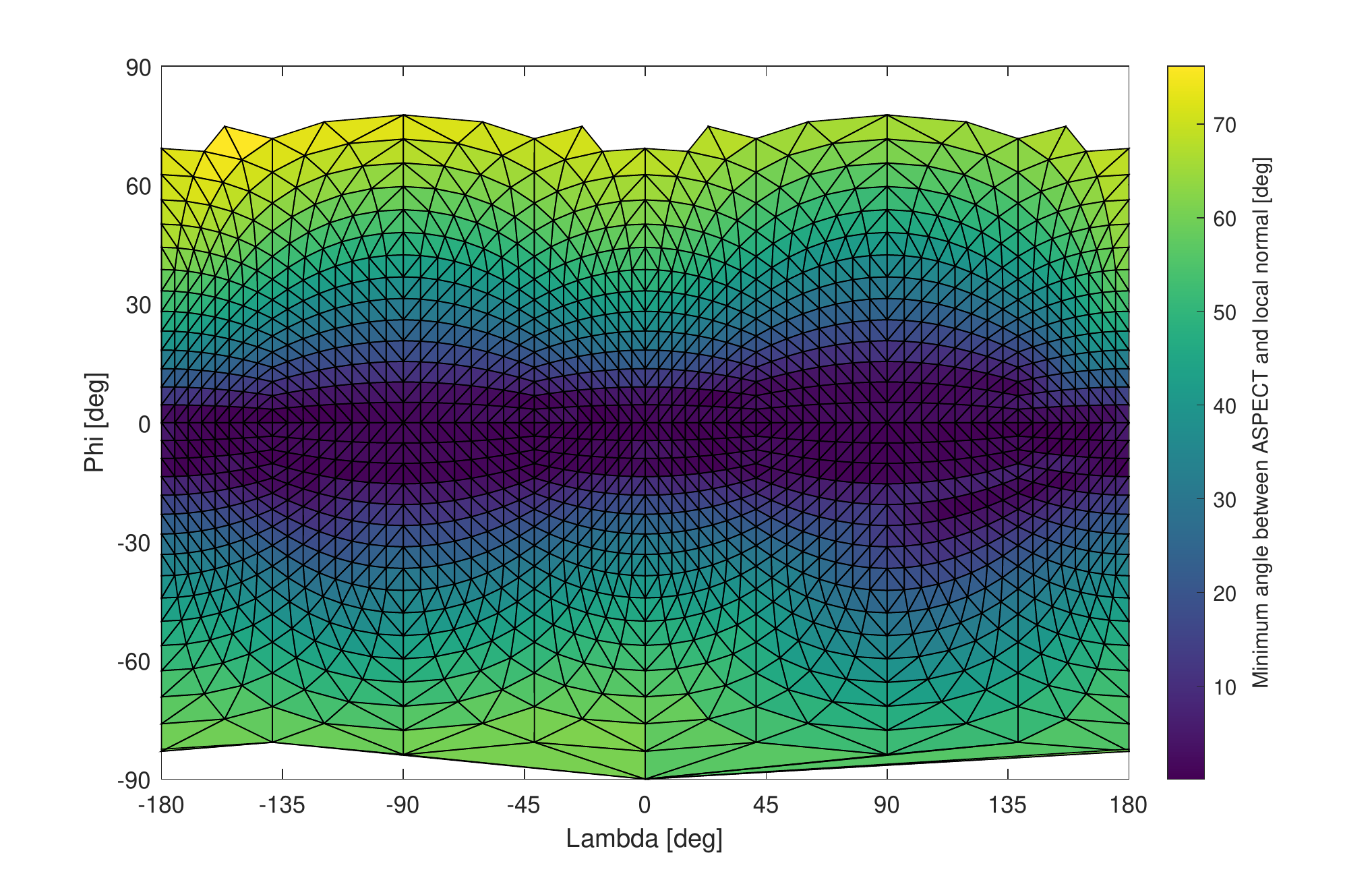}
       	\caption{ASPECT potential global coverage on Dimorphos. The map represents which faces of the shape model can be imaged by the payload given that the observation requirements are satisfied. The color of each face represents the minimum angle between the ASPECT line of sight and the face local normal during the science orbit cycle.}
       	\label{fig:coverage_D2}
    \end{figure}

\begin{figure}[p]
      	\centering
    	\includegraphics[width=0.99\textwidth]{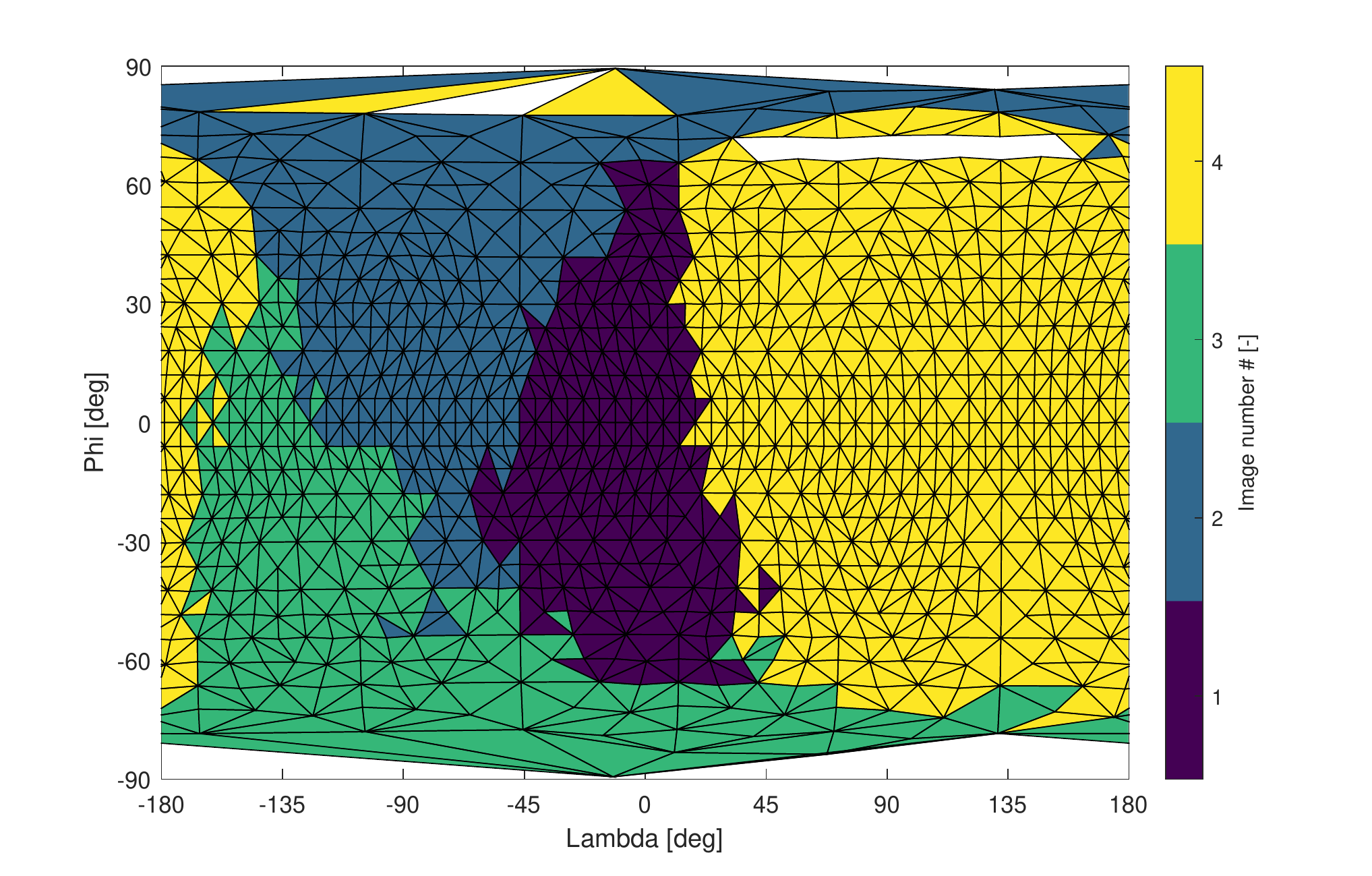}
       	\caption{Simulation of an image acquisition sequence to achieve global coverage on Didymos. The colors represent the areas of the shape model covered by each image in a sequential order. Global coverage on the visible faces can be achieved with 4 images.}
       	\label{fig:coverage_D1_4_images}
    \end{figure}
\begin{figure}[p]
      	\centering
    	\includegraphics[width=0.99\textwidth]{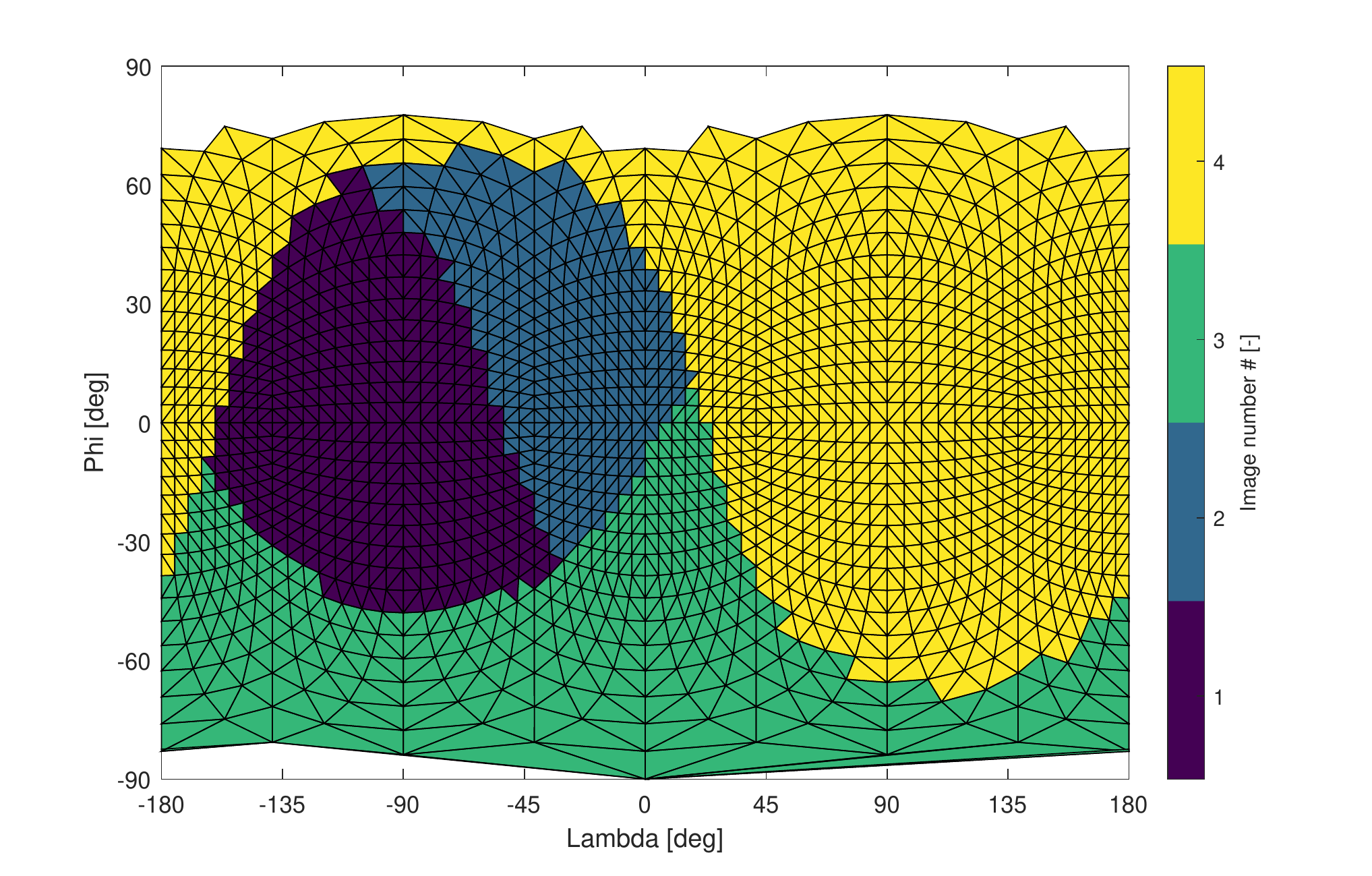}
       	\caption{Simulation of an image acquisition sequence to achieve global coverage on Dimorphos. The colors represent the areas of the shape model covered by each image in a sequential order. Global coverage on the visible faces can be achieved with 4 images.}
       	\label{fig:coverage_D2_4_images}
    \end{figure}

    Figure~\ref{fig:coverage_D1} and Figure~\ref{fig:coverage_D2} illustrate the faces that can be imaged during the potential observation periods. In bot cases global coverage can be achieved with the trajectory chosen. It is also estimated that a minimum number of 4~images timed at the proper epochs will be needed for each body to achieve global coverage. The areas covered by them is illustrated in Figure~\ref{fig:coverage_D1_4_images} and Figure~\ref{fig:coverage_D2_4_images}.

    This analysis is valid assuming ideal pointing and that the target body is entirely within the FOV of ASPECT. If the latter case is not true, especially in the case of Didymos, a mosaic strategy is proposed. As far as the scientific orbit is concerned, there seems to be plenty of opportunities in terms of observation time to acquire a higher number of images. The bottle-neck in this case might be given by the overall data budget, which is expected in the range of few Gbits.

    \begin{figure}[t]
      	\centering
    	\includegraphics[width=0.8\textwidth]{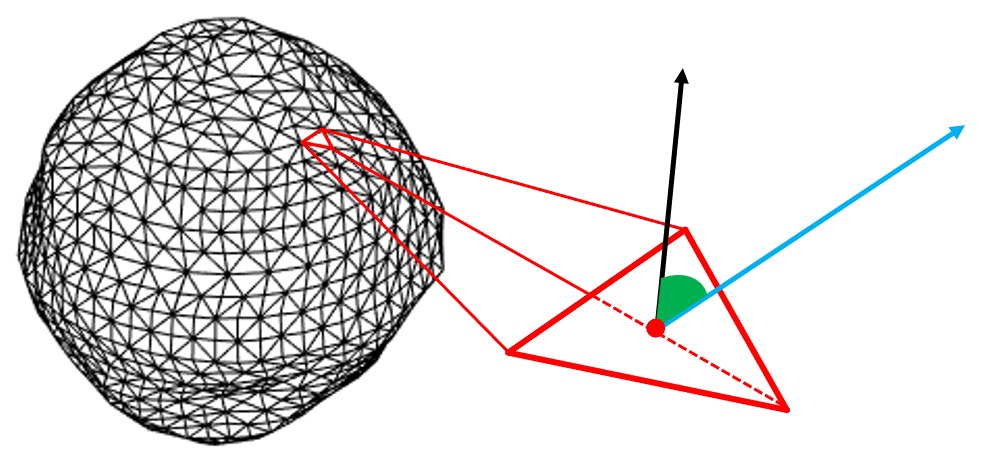}
       	\caption{Schematic of the angle used as metric to color the faces of the shape models in Figure \ref{fig:coverage_D1} and Figure \ref{fig:coverage_D2}. The angle is the one between the line of sight vector from the face center to ASPECT (cyan) and the local normal of each face (black).}
       	\label{fig:sketch_1_angle}
    \end{figure}

\section{Conclusion} 
\label{sec:6-conclusion}
    The paper presents the preliminary mission profile of Hera's Milani CubeSat, during its operational lifetime after release in the proximity of Didymos binary asteroid system. We provide a detailed analysis of the feasibility of such mission, given the constraints arising from Hera mission scenario. We discuss the challenges and driving design criteria in terms of mission analysis, trajectory design and GNC design. Finally, we provide a preliminary design solution, consistent with mission objectives and demonstrate its feasibility and suitability in meeting science and technological obejctives of the Milani CubeSat mission.
    
    Although the preliminary mission profile suggests feasibility against the imposed requirements and constraints, the initial results presented in this paper will be subject to changes as the Milani mission project evolves.

\section*{Acknowledgement}
    This work has been performed in response to ESA call AO/1-10258/20/NL/GLC: Hera Mission “Second CubeSat” Phase A/B C/D \& E1. Fabio Ferrari acknowledge funding from the European Union’s Horizon 2020 research and innovation programme under the Marie Skłodowska-Curie grant agreement No.\ 800060. Mattia Pugliatti and Francesco Topputo acknowledge funding from the European Union’s Horizon 2020 research and innovation programme under the Marie Skłodowska-Curie grant agreement no. 813644. The authors would like to acknowledge the support received by the whole Milani Team.

\bibliographystyle{apalike_long}

\bibliography{mybibfile}

\end{document}